\pdfoutput=1
\documentclass[5p,twocolumn,10pt,times]{elsarticle}
\usepackage{hyperref}
\usepackage{xcolor}
\usepackage{times}
\usepackage{graphicx}
\usepackage{amsmath,amssymb,amsthm}
\usepackage{array}
\usepackage{booktabs}
\usepackage{multirow}
\usepackage[normalem]{ulem}
\usepackage[ruled,linesnumbered]{algorithm2e}

\theoremstyle{plain}

\newcommand{\toolName}[1]{\textit{MidSurfNet}}

\begin{document}
\baselineskip11pt

\begin{frontmatter}

\title{\toolName{}: Learning Face Pairing for Mid-surface Abstraction of Thin-walled CAD Models}

\author[1]{Li Ye}
\ead{li-ye@zju.edu.cn}
\author[1]{Xinhang Zhou}
\ead{2xh@zju.edu.cn}
\author[1]{Xingyu Yang}
\ead{xy\_yang@zju.edu.cn}
\author[1]{Ruofeng Tong}
\ead{trf@zju.edu.cn}
\author[2]{Hailong Li}
\ead{lihailong@poissonsoft.com}
\author[1]{Peng Du}
\ead{dp@zju.edu.cn}
\author[1]{Min Tang\corref{mycorrespondingauthor}}
\cortext[mycorrespondingauthor]{Corresponding author.}
\ead{tang\_m@zju.edu.cn}

\address[1]{Zhejiang University, Hangzhou, 310007, China}
\address[2]{Shenzhen Poisson Software Co., Ltd., Shenzhen, 518129, China}

\begin{abstract}
Mid-surface abstraction is an important preprocessing step for finite element analysis of thin-walled CAD models, and face pairing is its central subproblem. Existing face-pairing methods rely primarily on handcrafted geometric criteria whose thresholds are difficult to tune when a model contains multiple local wall thicknesses; moreover, their variable-cardinality groupings depend on threshold settings and processing order, so the same model can yield inconsistent results. We present \toolName{}, a learning-based face-pairing method that couples a \textit{learned face-pair scorer} with a \textit{deterministic face-group composition}. The scorer evaluates every unordered face pair with two separately learned evidence streams---a geometry stream that combines continuous pairing criteria with a conditional shape correction, and an attributed-topology stream over the B-Rep face-adjacency graph. A pair-conditioned gate fuses the two streams, and independent per-pair decisions retain opposing-face support relations at a single operating threshold selected once on validation data, replacing the per-model geometric thresholds of rule-based pipelines rather than adding one. Under a connected-and-bipartite condition, the composition stage organizes the retained support relations into candidate variable-cardinality $m$-to-$n$ face groups, each independent of processing order for a fixed support graph and unique up to exchanging its two side labels. We also construct the MidSurf dataset, a benchmark of 1,575 manually annotated CAD models.
On the test set, \toolName{} attains a pair-level F1-Score of 87.32\%, 23.22 percentage points above the strongest rule-based baseline, and reaches an end-to-end Completion Rate of 75.42\%, including 61.90\% on the multi-wall-thickness category that the evaluated rule-based implementations do not support. We further demonstrate practical utility by generating mid-surfaces from the composed candidate face groups through an industrial mid-surface API and running finite element analyses on the resulting shell models.
\end{abstract}

\begin{keyword}
Mid-surface Abstraction, Face Pairing, Deep Learning, Thin-walled CAD Models, Benchmark Dataset
\end{keyword}

\end{frontmatter}

\begin{figure*}[!t]
    \centering
    \includegraphics[width=0.95\textwidth]{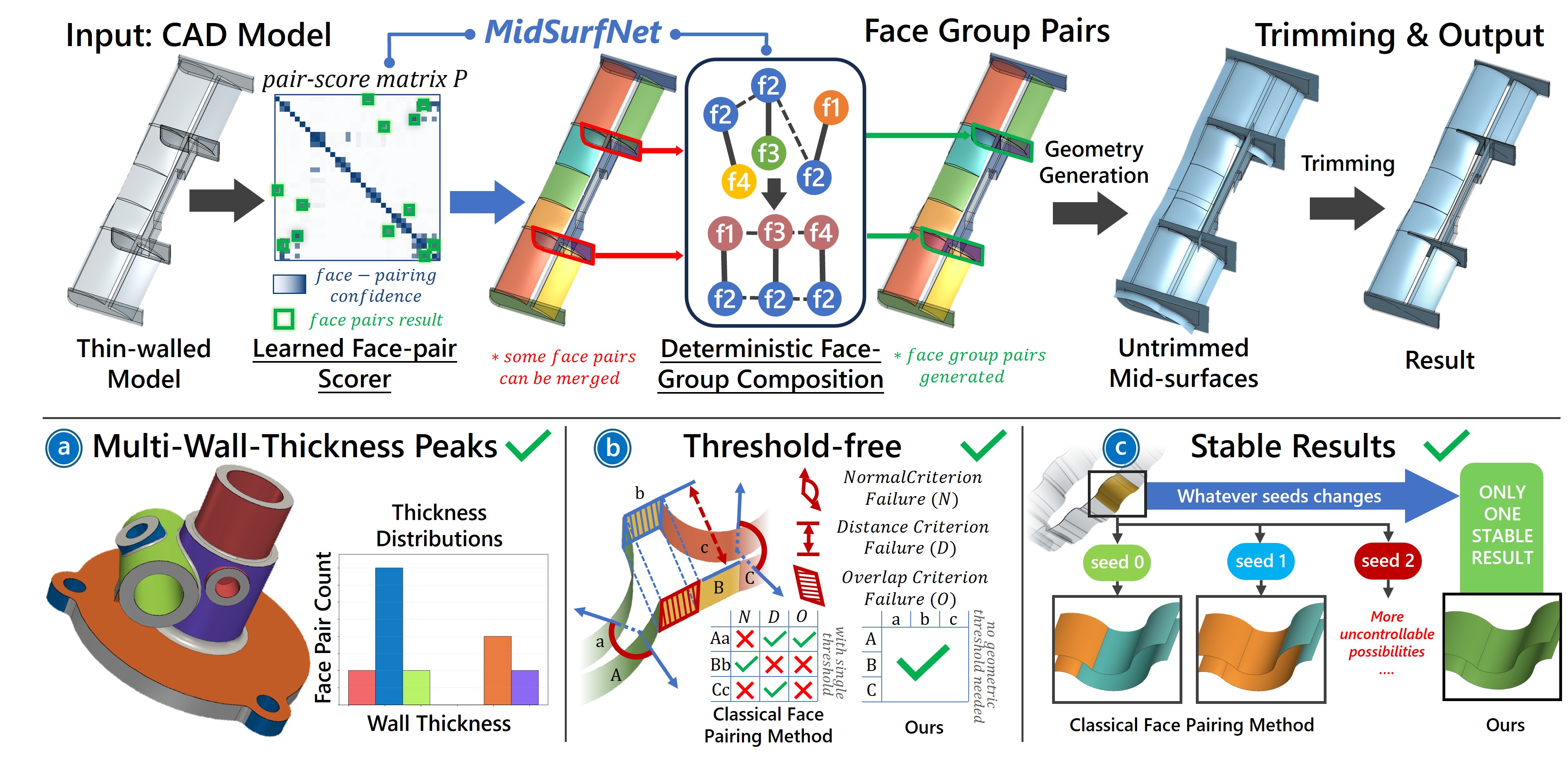}
    \vspace{-4mm}
    \caption{\toolName{} pipeline and the three challenges it addresses.
    \textbf{Top:} Given a thin-walled CAD model, the \textit{learned
    face-pair scorer} scores every unordered face pair into the pair-score
    matrix $\mathbf{P}$; the \textit{deterministic face-group composition}
    then organizes the retained support relations into candidate opposing
    face groups under the connected-and-bipartite condition, and a standard
    downstream step generates and trims the final mid-surfaces.
    \textbf{Bottom:} (a) several local wall thicknesses form separate peaks
    that no single distance cutoff covers; (b) classical criteria misjudge
    pairs under any single threshold, while the learned
    scorer requires no geometric threshold; (c) rule-based grouping varies from seed to seed, while the
    deterministic composition returns one stable result.}
    \label{fig:teaser}
\end{figure*}

\section{Introduction}
\label{sec:introduction}

Mid-surface abstraction is a foundational preprocessing task in computer-aided engineering (CAE). Thin-walled components are ubiquitous across aerospace, automotive, and mechanical engineering domains---from aircraft fuselages and automobile body panels to consumer electronics enclosures. By simplifying three-dimensional solid models into two-dimensional mid-surface representations, engineers can reduce computational degrees of freedom by an order of magnitude while maintaining plate/shell analysis fidelity, thereby significantly improving finite element analysis efficiency~\cite{rezayat1996midsurface}.

Existing mid-surface abstraction methods fall into three primary categories: Medial Axis Transform (MAT), Chordal Axis Transform (CAT), and face pairing. MAT generates skeletal structures by computing the locus of maximal inscribed spheres, but its topological complexity necessitates extensive manual pruning. CAT converts thin-walled solids into single-layer tetrahedral meshes, but struggles to maintain geometric quality in complex junction regions. In contrast, face-pairing methods identify opposing wall faces and construct intervening sheets, preserving boundary-representation (B-Rep) modeling intent; this has made them the de facto CAD/CAE standard.

However, current face-pairing methods face several fundamental limitations. First, \textit{multi-wall-thickness failures}: when one model contains several distinct local wall thicknesses, no distance cutoff is simultaneously correct for all of its walls, so any global setting misses some true pairs or admits spurious ones (Fig.~\ref{fig:teaser}-a).
Second, \textit{threshold dependence}: classical criteria accept a candidate pair only when distance, normal deviation, and projected overlap pass prescribed thresholds; these thresholds must be tuned per model family, and a setting that suits one geometry class transfers poorly to another (Fig.~\ref{fig:teaser}-b).
Third, \textit{unstable variable-cardinality grouping}: when one side of a wall is fragmented into several B-Rep faces, rule-based pipelines assemble opposing face groups through greedy, order-dependent merging, so small threshold perturbations or a different processing order can produce inconsistent groupings for the same model (Fig.~\ref{fig:teaser}-c).

Recent advances in deep learning for CAD geometry processing present opportunities to overcome these limitations. Neural networks have demonstrated remarkable capabilities in learning complex geometric patterns from data, as evidenced by BRepNet~\cite{lambourne2021brepnet} and UV-Net~\cite{jayaraman2021uvnet} for B-Rep understanding. This suggests that supervised learning on annotated support relations can capture local opposition evidence beyond explicit geometric rules, while a deterministic graph composition can assemble the retained relations into a candidate group partition whenever they connect all faces of a wall and remain two-sided (formalized in Sec.~\ref{sec:problem_formulation}).

We introduce \toolName{}, to our knowledge, the first learning-based face-pairing method for mid-surface abstraction, addressing the aforementioned challenges. As illustrated in Fig.~\ref{fig:teaser}, given an input B-Rep model, the \textit{learned face-pair scorer} evaluates every unordered face pair through separately learned geometry and attributed-topology streams, fused by a pair-conditioned gate. Independent per-pair decisions then retain opposing-face support relations at one operating threshold selected on validation data; because no one-to-one capacity constraint is imposed, one face may retain relations to several others. The \textit{deterministic face-group composition} finally organizes a connected and bipartite union of overlapping one-to-many support patterns into candidate variable-cardinality $m$-to-$n$ face groups, each with its two side sets unique up to exchanging their labels, independent of processing order. The candidate groups remain editable when application-specific semantics are required and are then passed to a standard downstream mid-surface construction step. Our contributions are:

\begin{itemize}
    \item \textbf{\toolName{}}: A \textit{learned face-pair scorer} that keeps the classical distance, normal, and overlap criteria as continuous evidence rather than acceptance tests, conditionally corrects this physical baseline with compact shape evidence, and arbitrates between geometric and attributed-topology evidence per pair; a \textit{deterministic face-group composition} then composes group structure rather than predicting it, with no additional geometric threshold.

    \item \textbf{MidSurf dataset}: A large-scale benchmark for mid-surface abstraction comprising 1,575 thin-walled CAD models with manually annotated one-to-one opposing-face support relations, spanning constant-, variable-, and multi-wall-thickness categories, with an inter-annotator agreement of $\kappa=0.91$ and a deduplicated, stratified split protocol.
\end{itemize}

\section{Related Work}
\label{sec:related_work}

We discuss three lines of related work: classical mid-surface abstraction, which defines the problem and its rule-based state of the art; learning on B-Rep models, which underpins attributed-topology encoding and the pairwise task itself; and correcting and fusing evidence sources, which motivates our criteria-grounded shape correction (Sec.~\ref{sec:criterion_residual}) and pair-conditioned fusion (Sec.~\ref{sec:fusion_decision}).

\subsection{Mid-surface Abstraction Methods}
\label{sec:related_midsurf}

Mid-surface abstraction transforms thin-walled solid models into dimensionally reduced representations for finite element analysis~\cite{thakur2009survey}. Three main approaches have been developed: MAT, CAT, and face pairing.

MAT, introduced by Blum~\cite{blum1967transformation}, computes the locus of centers of maximal inscribed spheres. While theoretically elegant, MAT is highly sensitive to boundary perturbations and generates numerous small branches that require extensive pruning~\cite{sherbrooke1996algorithm,amenta2001power,dey2003approximate}; recent formulations improve its robustness and geometric quality~\cite{wang2022mat,kong2024quasi}, yet substantial post-processing is still needed to obtain analysis-ready sheets. CAT methods~\cite{prasad1997morphological,quadros2002hexlayer} convert thin-walled solids into single-layer tetrahedral meshes and extract chordal sheets, but struggle to maintain geometric quality at complex junction regions. Learned skeletonization exists for point clouds~\cite{lin2021point2skeleton}, but no learned method targets B-Rep face pairing.

Face pairing has emerged as the mainstream approach in CAD/CAE practice. The Face Adjacency Graph formulation~\cite{rezayat1996midsurface} matches opposing faces using distance, normal, and overlap criteria. Subsequent work improved topological integrity through boundary extension~\cite{sheen2010transformation}, normal-projection pairing with mid-surface fitting heuristics~\cite{zhu2016mid}, and divide-and-conquer decomposition~\cite{woo2014divide}; thick--thin decomposition~\cite{sun2017decomposing,nolan2019analysis} separates regions amenable to pairing from residual solid volumes. MidSurfer~\cite{ye2026midsurfer} introduced an automated pipeline for variable thin-walled models, and commercial systems expose mid-surface operations built on the same criteria~\cite{Parasolid}. These pipelines accept candidate pairs through threshold-based geometric tests and require additional handling for multiple local wall thicknesses; moreover, because opposing face groups are assembled by rule application, the recovered variable-cardinality groupings can depend on threshold settings and processing order. Our learning-based approach instead learns support evidence without prescribed geometric acceptance thresholds and composes the resulting support graph without additional geometry-based completion rules.

\subsection{Deep Learning on B-Rep Models}
\label{sec:related_dl_cad}

A growing body of work learns directly on the B-Rep data structure: UV-Net~\cite{jayaraman2021uvnet} samples each face in its parametric domain and couples per-face convolutional features with a face-adjacency graph network, BRepNet~\cite{lambourne2021brepnet} defines convolution kernels over coedges, and the B-Rep Transformer (BRT)~\cite{zou2025bringing} encodes exact surface geometry with attention over wire--face hierarchies. These general-purpose encoders entangle all geometric and topological evidence in one representation. Our design instead controls where learning may act inside the learned face-pair scorer: the geometry stream preserves the physically interpretable pair criteria and adds compact shape evidence only as a conditional correction (Secs.~\ref{sec:geometry_stream} and~\ref{sec:criterion_residual}); graph learning is confined to the separate attributed-topology stream, where opposing wall faces generally do not share an edge but are connected by short paths, so plain message passing~\cite{kipf2017semi,velivckovic2018graph} over the attributed B-Rep face-adjacency graph suffices, with longer-range structure injected only through a per-pair all-shortest-route context rather than global attention~\cite{ying2021transformers} (Sec.~\ref{sec:topology_stream}).

Closest to our setting are learned relation tasks on B-Reps from assembly modeling: AutoMate~\cite{jones2021automate} proposes mates between parts, and JoinABLe~\cite{willis2022joinable} predicts joints connecting two parts; learned matching in vision shares the same structure, resolving two descriptor sets into an assignment through optimal transport~\cite{sarlin2020superglue}, dense attention~\cite{sun2021loftr}, or graph matching~\cite{wang2019learning}. All of these establish correspondences \emph{between} two entity sets. Face pairing is instead a \emph{within-part} relation task: one wall side may consist of several faces, many faces must remain unmatched as lateral faces (Sec.~\ref{sec:problem_formulation}), and no assignment or capacity constraint applies. We therefore decide every unordered pair independently (Sec.~\ref{sec:fusion_decision}) and form candidate group structure through a separate deterministic composition (Sec.~\ref{sec:group_composition}).

\subsection{Correcting and Fusing Evidence Sources}
\label{sec:related_hybrid}

Two mechanisms at the core of our method---learning a bounded correction to a physically grounded baseline, and weighting complementary evidence sources for each input---have long histories outside CAD.

Learning a correction on top of an existing predictor is a classical idea. Gradient boosting fits every new model to the residual of the current ensemble~\cite{friedman2001greedy}, and probability calibration adjusts a trained scorer without changing what it ranks first~\cite{platt1999probabilistic,guo2017calibration}. ControlNet~\cite{zhang2023adding} showed that an additional conditioning branch can be introduced through zero-initialized layers so that it does not perturb the current predictor at initialization. Our conditional shape correction (Sec.~\ref{sec:criterion_residual}) follows the same identity-start principle but assigns the physical roles differently: distance, normal, and overlap measurements define the learned criterion baseline, while compact shape evidence contributes only a bounded, pair-dependent correction. Thus, quantities used as hard acceptance tests in classical pipelines remain continuous evidence, and the added shape branch initially leaves the criterion score unchanged.

How to weight several sources of evidence is an equally established question. Two-stream networks process complementary modalities separately and fuse them late~\cite{simonyan2014two}; mixtures of experts learn input-dependent gates over specialist predictors~\cite{jacobs1991adaptive,shazeer2017outrageously}; and uncertainty-aware models condition prediction on estimated uncertainty~\cite{kendall2017uncertainties,kendall2018multi}. Our pair-conditioned fusion (Sec.~\ref{sec:fusion_decision}) operates at the granularity of one face pair: independent geometry and topology logits are merged by a convex gate conditioned on both logits, their disagreement, and a latent topology-conditioning signal. This signal has no separate reliability target, and the geometry-dominated initialization merely gives the topology branch a conservative starting influence rather than certifying its trustworthiness.

\section{Preliminaries}
\label{sec:preliminaries}

\subsection{Problem Formulation}
\label{sec:problem_formulation}

Given a thin-walled B-Rep model $\mathcal{M}$ with faces $\mathcal{F} = \{f_1, f_2, \ldots, f_N\}$, an \textit{opposing face group} $(\mathcal{A}_k,\mathcal{B}_k)$ is the pair of face sets that jointly bound one physical wall; with $m=|\mathcal{A}_k|$ and $n=|\mathcal{B}_k|$, the cases $m=n=1$, $\min(m,n)=1<\max(m,n)$, and $m,n>1$ define one-to-one, one-to-many, and many-to-many groups, and exchanging the two sets does not change the group. An \textit{opposing-face support relation} is an unordered pair $(f_i,f_j)$ whose corresponding regions directly oppose one another across part of the same wall. A one-to-one group fixes exactly one such relation, but a variable-cardinality group does not prescribe a unique pairwise target: fragmentation is local, and a face need not directly oppose every fragment on the other side. Any side-consistent set of support relations that spans $\mathcal{A}_k\cup\mathcal{B}_k$ and makes it connected is therefore a \textit{witness support graph} for the group.

Every paired face belongs to at most one wall group (the face-disjoint output contract), and faces belonging to no group are \textit{lateral faces}. The learned task predicts support relations; a subsequent graph composition forms candidate group partitions, and exact recovery is conditional on the predictions forming a valid witness graph for each group with no edge leaving it (Sec.~\ref{sec:group_composition}).

Face pairing is the central subproblem of mid-surface abstraction. The mid-surface $\textit{MS}$ of an opposing face group lies between the two sides, with its position set by an offset parameter $\alpha \in [0,1]$~\cite{rezayat1996midsurface}:
\begin{equation}
    \forall P \in \textit{MS}, \quad (1-\alpha)\,\text{Dist}(P, \mathcal{A}_k) = \alpha\,\text{Dist}(P, \mathcal{B}_k)
    \label{eq:midsurface_def}
\end{equation}
where $\text{Dist}(P, \mathcal{S})$ is the minimum Euclidean distance from point $P$ to the faces in set $\mathcal{S}$; as $\alpha \to 0$ the surface approaches $\mathcal{A}_k$, and as $\alpha \to 1$ it approaches $\mathcal{B}_k$ (Fig.~\ref{fig:Preliminaries}-b). The standard case is the centered one, $\alpha = \tfrac{1}{2}$, where \eqref{eq:midsurface_def} reduces to equal distances to the two sides; this centered mid-surface is the default target in CAE practice and the one used throughout this paper.

\subsection{Face Pairing Criteria}
\begin{figure}[t]
    \centering
    \includegraphics[width=\linewidth]{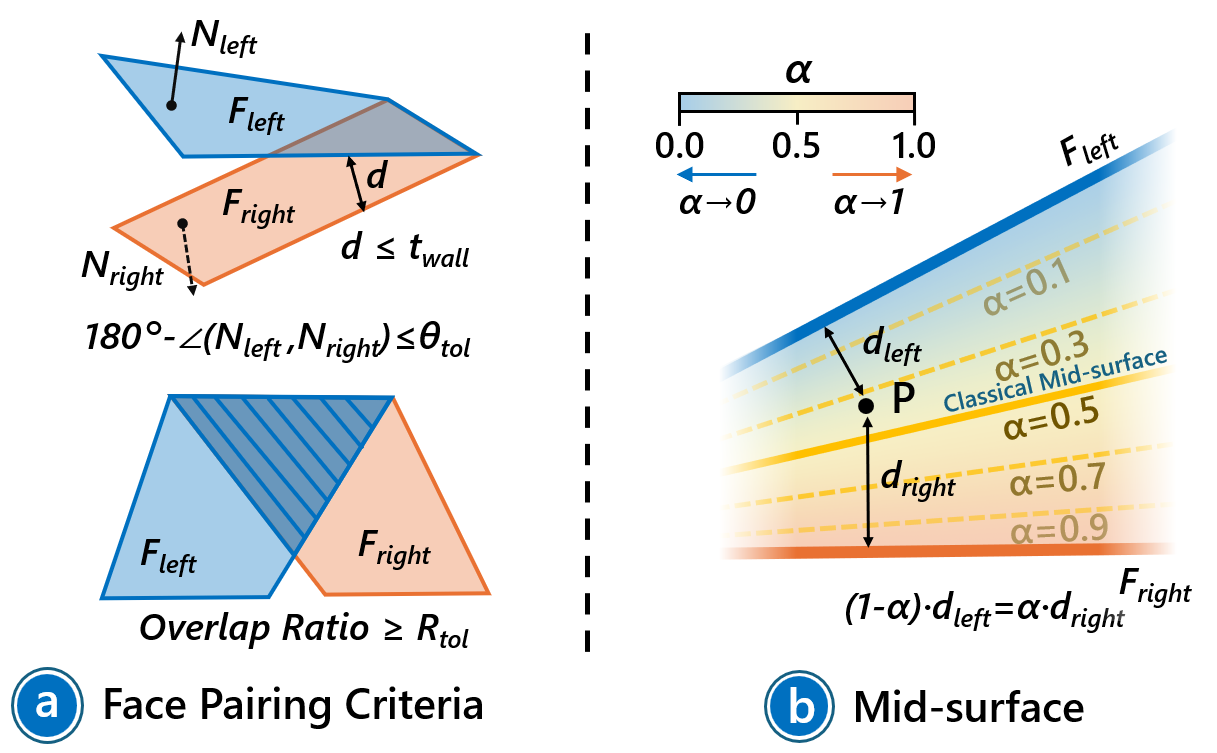}
    \vspace{-6mm}
    \caption{Foundations of face pairing. (a) Classical pairing criteria: distance, normal, and overlap. (b) The mid-surface of an opposing face group: the offset parameter $\alpha$ sets its position between the two sides, and $\alpha = \tfrac{1}{2}$ gives the centered mid-surface used in this paper.}
    \vspace{-5mm}
    \label{fig:Preliminaries}
\end{figure}

\label{sec:fp_criteria}

As shown in Fig.~\ref{fig:Preliminaries}-a, traditional face-pairing methods evaluate candidate pairs based on three geometric criteria:

\textbf{Distance Criterion.} The distance between the two faces of a candidate pair $(f_i, f_j)$ must not exceed the wall thickness threshold:
\begin{equation}
    d(f_i, f_j) \leq t_{\text{wall}}
    \label{eq:distance_criterion}
\end{equation}
where $d(f_i, f_j)$ is the minimum Euclidean distance between the faces.

\textbf{Normal Criterion.} The outward normals of paired faces must be approximately antiparallel:
\begin{equation}
    180^\circ - \angle(\boldsymbol{\nu}_i, \boldsymbol{\nu}_j) \leq \tau_{\mathrm{n}}
    \label{eq:normal_criterion}
\end{equation}
where $\boldsymbol{\nu}$ denotes a face normal and $\tau_{\mathrm{n}}$ is an angular tolerance (typically $10^\circ$--$30^\circ$). We write $n_{ij}=\cos\angle(\boldsymbol{\nu}_i,\boldsymbol{\nu}_j)$ for the normal cosine of a pair.

\textbf{Overlap Criterion.} Overlap is directional: let $o_{ij}$ be the fraction of $f_i$ that projects onto $f_j$ along the normal direction of $f_i$, so $o_{ij}\neq o_{ji}$ in general. Both directions must exceed a threshold:
\begin{equation}
    \min\!\left(o_{ij},\, o_{ji}\right) \geq \tau_{\mathrm{o}}
    \label{eq:overlap_criterion}
\end{equation}
where $\tau_{\mathrm{o}}$ is typically set between 50\% and 80\%.

These criteria form the foundation of rule-based face pairing; Sec.~\ref{sec:geometry_stream} retains the same three measurements as continuous inputs to a learned criterion baseline instead of pass--fail tests.

\begin{figure*}[t]
    \centering
    \includegraphics[width=\linewidth]{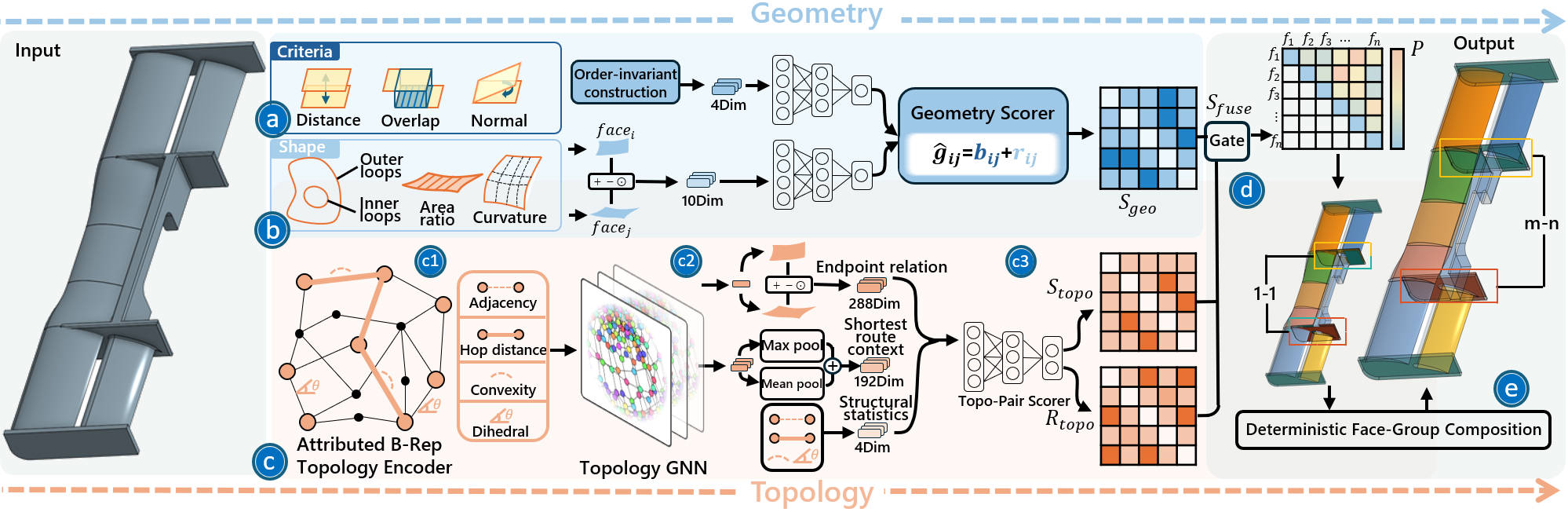}
    \caption{The \toolName{} pipeline. The \textit{learned face-pair scorer}
    (a--d) scores every unordered face pair with two separately learned
    streams: the geometry stream combines criterion-baseline scoring (a) with
    a conditional shape correction (b), and the attributed-topology stream (c)---face attributes (c1), topology
    GNN (c2), and Topo-Pair Scorer (c3)---supplies a topology logit and a
    conditioning signal; a pair-conditioned
    gate fuses the two streams and a single threshold retains opposing-face
    support relations (d). The \textit{deterministic face-group composition}
    (e) assembles the retained relations into candidate variable-cardinality
    face groups.}
    \label{fig:fp_architecture}
\end{figure*}
\section{Method}
\label{sec:method}

\subsection{Method Overview}
\label{sec:method_overview}

Given a thin-walled B-Rep model with faces
$\mathcal{F}=\{f_1,\dots,f_N\}$ (Sec.~\ref{sec:problem_formulation}),
\toolName{} proceeds in three stages; the \textit{learned face-pair scorer} that
implements the first two is detailed in
Fig.~\ref{fig:fp_architecture}:

\begin{enumerate}
    \item \textbf{Dual-stream face-pair scoring} (Sec.~\ref{sec:neural_face_pairing}): the geometry stream (upper lane of Fig.~\ref{fig:fp_architecture}) produces a geometry-logit matrix $\mathbf{S}_{\mathrm{geo}}$, while the topology stream (lower lane) produces a topology-logit matrix $\mathbf{S}_{\mathrm{topo}}$ and a topology-conditioning matrix $\mathbf{R}_{\mathrm{topo}}$ over the same unordered face pairs.
    \item \textbf{Pair-conditioned fusion and support-relation decision} (Sec.~\ref{sec:fusion_decision}): the Pair-Conditioned Gate combines the two branch logits into the fused-logit matrix $\mathbf{S}_{\mathrm{fuse}}$. A sigmoid and a single validation-selected threshold $\theta$ then decide which opposing-face support relations are retained (Fig.~\ref{fig:fp_architecture}-d); these decisions are independent.
    \item \textbf{Deterministic face-group composition} (Sec.~\ref{sec:group_composition}): Algorithm~\ref{alg:group_composition} decomposes the thresholded support graph into connected components and two-colors each bipartite component (Fig.~\ref{fig:fp_architecture}-e). Overlapping one-to-many relations thereby compose into candidate variable-cardinality $m$-to-$n$ groups without learned parameters. A non-bipartite component violates the composition contract and is treated as a composition failure rather than a valid group.
\end{enumerate}

The composed candidate face groups are subsequently passed to the industrial mid-surface API of Siemens NX, which constructs and trims the final mid-surfaces. This downstream operation follows standard practice in mid-surface abstraction~\cite{woo2014divide,ye2026midsurfer} and is not a contribution of this work.

Four design principles organize the learned scorer. \textbf{(P1) Complete,
order-invariant support scoring:} all $N(N-1)/2$ unordered pairs are evaluated
without candidate pruning, and every representation is invariant to exchanging
the two faces. Completeness refers to the evaluated pair domain, not to a
complete-bipartite output: the retained support graph need not contain every
cross-side combination, and a face
may have multiple incident support edges. \textbf{(P2) Criteria-grounded shape correction:}
the classical pairing criteria define a continuous physical baseline, while
shape supplies only a bounded correction conditioned on the current pair.
\textbf{(P3) Topology as context, not as a correspondence constraint:} the
attributed B-Rep graph describes structural roles and connections without
restricting which faces may be paired. \textbf{(P4) Pair-conditioned fusion:}
a learned gate selects the geometry--topology mixture separately for each pair rather
than using one global coefficient. Fig.~\ref{fig:fp_architecture}
instantiates this division of labor: the Geometry Scorer implements the
criterion--shape hierarchy, the attributed-topology stream remains separate,
and the Pair-Conditioned Gate is their first interaction. These principles
also organize the ablations in
Sec.~\ref{sec:ablation}; individual descriptors and network layers are treated
as implementation details.

\subsection{Dual-Stream Face-Pair Scoring}
\label{sec:neural_face_pairing}

Face pairing is a complete relation-prediction problem rather than a sequence
of candidate filters. For every unordered pair, geometry must determine whether
the measured surfaces can physically form two wall sides, while topology must
determine whether they occupy compatible roles in the surrounding B-Rep.
Neither question subsumes the other. As illustrated in
Fig.~\ref{fig:fp_architecture}, architecturally separate geometry and
attributed-topology streams score the same complete pair domain; the geometry
stream further separates a criterion baseline from conditional shape
correction. They remain separate until the Pair-Conditioned Gate of
Sec.~\ref{sec:fusion_decision}, so topology can resolve geometric ambiguity
without becoming a candidate constraint.

\subsubsection{Geometric Criteria and Criterion-Baseline Scoring}
\label{sec:geometry_stream}

Opposing sides of a thin wall have a characteristic geometric relation:
their separation reflects the local wall thickness, their outward normals
point in opposite directions, and their projected regions overlap. These
are the same cues measured by the classical criteria of
Sec.~\ref{sec:fp_criteria} (Fig.~\ref{fig:fp_architecture}-a). Their limitation lies not in the measurements, but
in turning each one into a fixed pass--fail test before considering their joint
evidence. Because wall thickness and face fragmentation vary within a B-Rep
model, a fixed set of cutoffs need not define a stable acceptance region across
these variations. We instead retain the criteria as continuous inputs and learn
their joint relation to face opposition.

For every unordered pair $i<j$, we canonically order the two directional
overlap ratios and construct the four-dimensional, order-invariant criterion
vector
\begin{equation}
  \mathbf{c}_{ij}
  =[\,d_{ij},\;n_{ij},\;
  \min(o_{ij},o_{ji}),\;\max(o_{ij},o_{ji})\,]
  =\mathbf{c}_{ji}\in\mathbb{R}^{4},
\label{eq:continuous_criterion_vector}
\end{equation}
where $d_{ij}=d(f_i,f_j)/L$ is the distance normalized by the model
bounding-box diagonal $L$, $n_{ij}$ is the normal cosine, and
$o_{ij},o_{ji}$ are the directional overlap ratios. Their minimum and maximum
form a canonical representation of the unordered overlap pair, retaining the
magnitude of its directional imbalance while discarding arbitrary face order:
the minimum records the overlap supported in both directions, whereas the
maximum preserves one-sided containment caused by unequal face splitting. A multi-layer perceptron (MLP)
$\Phi_{\mathrm{crit}}$ maps this vector to the criterion-baseline logit
\begin{equation}
  b_{ij}=\Phi_{\mathrm{crit}}(\mathbf{c}_{ij})=b_{ji}.
  \label{eq:criterion_base_logit}
\end{equation}
No element of $\mathbf{c}_{ij}$ is thresholded, and all $N(N-1)/2$ pairs are
evaluated. Thus, $b_{ij}$ is a complete, physically grounded criterion baseline
rather than the output of a hand-crafted filter; Sec.~\ref{sec:shape_correction}
refines this baseline without changing the pair domain.

\subsubsection{Compact Shape Evidence and Conditional Shape Correction}
\label{sec:shape_correction}
\label{sec:criterion_residual}

The criterion-baseline logit $b_{ij}$ captures relative placement but not
trimmed shape. Opposing faces may share curvature and loop layouts, yet repeated
features can make unrelated faces look similar, while trimming or splitting can
make a true pair dissimilar. Shape compatibility is thus neither sufficient nor
necessary for opposition; we use it only as a conditional correction to
$b_{ij}$, not as an independent acceptance rule.

For each unordered pair, the exchange-invariant shape vector
$\mathbf{x}_{ij}=\mathbf{x}_{ji}\in\mathbb{R}^{10}$ collects seven
comparisons---of positive and negative Gaussian-curvature fractions,
outer-boundary and edge/corner patterns, inner loops and their count, and
relative face area---and three availability indicators that prevent missing
measurements from being interpreted as agreement
(Fig.~\ref{fig:fp_architecture}-b). A shape head maps it to the raw logit $a_{ij}$ and bounded signed
evidence
\begin{equation}
  a_{ij}=\Phi_{\mathrm{shape}}(\mathbf{x}_{ij}),
  \qquad
  q_{ij}=\tanh\!\left(\tfrac{1}{2}a_{ij}\right)\in(-1,1),
  \label{eq:signed_shape_evidence}
\end{equation}
where $q_{ij}>0$ supports the pair and $q_{ij}<0$ contradicts it. Because
agreement and disagreement fail for different reasons, we predict separate,
pair-conditioned amplitudes for the two signs:
\begin{equation}
  (w^+_{ij},w^-_{ij})=
  \sigma\!\left(\Phi_{\mathrm{cond}}(
  [\,\mathbf{c}_{ij}\,\|\,\sigma(b_{ij})\,\|\,
  |q_{ij}|\,\|\,\mathbf{x}_{ij}\,])\right)\in(0,1)^2.
  \label{eq:conditional_shape_weights}
\end{equation}
Here $\sigma$ is the logistic function and $[\cdot\,\|\,\cdot]$ denotes
concatenation. The four inputs expose the physical relation, current baseline
state, evidence magnitude, and shape pattern and availability, respectively.

Let $v^{\mathrm{s}}_{ij},v^{\mathrm{o}}_{ij},v^{\mathrm{i}}_{ij}\in\{0,1\}$
indicate that surface, outer-boundary, and inner-loop evidence---the three
shape-evidence families---is available for both faces. With
$v_{ij}=\max(v^{\mathrm{s}}_{ij},v^{\mathrm{o}}_{ij},v^{\mathrm{i}}_{ij})$,
the correction and geometry logit are
\begin{equation}
  r_{ij}=v_{ij}
  \begin{cases}
    w^+_{ij}q_{ij}, & q_{ij}\geq0,\\
    w^-_{ij}q_{ij}, & q_{ij}<0,
  \end{cases}
  \qquad
  \hat{g}_{ij}=b_{ij}+r_{ij}.
  \label{eq:corrected_logit}
\end{equation}
If no shape family is available, $r_{ij}=0$ and the geometry stream falls back
to $b_{ij}$. Otherwise, $|q_{ij}|<1$ and $0<w^{\pm}_{ij}<1$ ensure
$|r_{ij}|<1$: shape remains a bounded additive correction rather than an
unrestricted alternative score. The resulting $\hat{g}_{ij}$ is the geometry logit used by the fusion
stage; collected over all pairs, these form
$\mathbf{S}_{\mathrm{geo}}=[\hat{g}_{ij}]$ in
Fig.~\ref{fig:fp_architecture}.

\subsubsection{Attributed B-Rep Topology Encoder and Topo-Pair Scoring}
\label{sec:topology_stream}

A face pair is judged geometrically from a local relation, whereas a thin wall
belongs to the boundary structure of an entire solid. Geometry therefore asks
whether two surfaces can form opposite wall sides; topology asks whether they
occupy compatible structural roles. The B-Rep face-adjacency graph---one node per face, one edge wherever two
faces share a B-Rep edge---supplies context that local geometry lacks, but it
cannot serve as the matching graph because opposing wall faces generally do
not share an edge. We therefore use topology to
contextualize every unordered pair, not to prune non-adjacent pairs.

The attributed B-Rep topology encoder (Fig.~\ref{fig:fp_architecture}-c)
uses adjacency and topological-distance statistics to describe graph position,
and convexity and dihedral attributes to describe local boundary folding
(Fig.~\ref{fig:fp_architecture}-c1). A Topology GNN with three message-passing layers
(Fig.~\ref{fig:fp_architecture}-c2) maps
them to structural face tokens $\mathbf{t}_i\in\mathbb{R}^{96}$ without
receiving criterion or shape features.
Symmetric combinations of the two endpoint tokens (the Endpoint Relation
terms of Eq.~\eqref{eq:topology_pair_descriptor}) relate a pair directly;
they cannot, however, describe how two non-adjacent neighborhoods are
connected. Face splitting and openings may also create
several equally short routes, so selecting one route would introduce an
arbitrary traversal choice. The all-shortest-route context instead
collects the union of interior faces lying on any shortest route,
\begin{equation}
  \mathcal{P}_{ij}=\{\,k\notin\{i,j\}:D_{ik}+D_{kj}=D_{ij}\,\},
  \label{eq:all_geodesic_faces}
\end{equation}
where $D_{ij}$ is the shortest-path distance between $f_i$ and $f_j$ on the
face-adjacency graph, with faces in different connected components assigned
a sentinel distance of $N$ so that their route context is empty; the tokens
of $\mathcal{P}_{ij}$ are then summarized by mean and maximum pooling (zero
vectors when $\mathcal{P}_{ij}$ is empty, as for adjacent faces or
cross-component pairs). The
resulting context is independent of which equally short route is found first.

The topo-pair feature concatenates symmetric endpoint relations, pooled
all-shortest-route context, and structural statistics:
\begin{equation}
\begin{aligned}
  \mathbf{z}_{ij}=[\,&\mathbf{t}_i+\mathbf{t}_j\,\|\,
  |\mathbf{t}_i-\mathbf{t}_j|\,\|\,
  \mathbf{t}_i\odot\mathbf{t}_j\,\|\\
  &\operatorname{mean}_{k\in\mathcal{P}_{ij}}\mathbf{t}_k\,\|\,
  \operatorname{max}_{k\in\mathcal{P}_{ij}}\mathbf{t}_k\,\|\,
  \boldsymbol{\eta}_{ij}\,]\in\mathbb{R}^{484},
\end{aligned}
\label{eq:topology_pair_descriptor}
\end{equation}
where $\odot$ is the element-wise product and
$\boldsymbol{\eta}_{ij}\in\mathbb{R}^{4}$ records graph distance, neighbor
overlap, route coverage, and degree balance. The Topo-Pair Scorer (Fig.~\ref{fig:fp_architecture}-c3) maps $\mathbf{z}_{ij}$ to a
topology-compatibility logit $u_{ij}$
and a topology-conditioning signal $\rho_{ij}\in(0,1)$. Collecting them over
all pairs gives $\mathbf{S}_{\mathrm{topo}}=[u_{ij}]$ and
$\mathbf{R}_{\mathrm{topo}}=[\rho_{ij}]$ in
Fig.~\ref{fig:fp_architecture}. Every term is
invariant to exchanging $i$ and $j$. As in the geometry stream, all
$N(N{-}1)/2$ pairs are scored, so the complete pair domain needed for
non-local opposition is preserved.

\subsection{Pair-Conditioned Fusion and Support-Relation Decision}
\label{sec:fusion_decision}

The two streams tend to disagree on difficult pairs: geometry is decisive
for regular parallel walls but ambiguous near fillets or fragmented faces,
whereas topology is informative in a well-connected shell but weaker for
isolated or heavily trimmed faces. A global coefficient would average over
these regimes. Input-dependent gates provide a general way to arbitrate
specialized predictors~\cite{jacobs1991adaptive,shazeer2017outrageously}; we
apply this principle to each face pair. The Pair-Conditioned Gate receives the
two branch logits and the topology-conditioning signal, while explicitly
exposing their disagreement and sigmoid-transformed scores:
\begin{equation}
\begin{aligned}
  \mathbf{m}_{ij}&=[\,\hat{g}_{ij},\,
  u_{ij},\, |\hat{g}_{ij}-u_{ij}|,\,
  \sigma(\hat{g}_{ij}),\,\sigma(u_{ij}),\,\rho_{ij}\,]
  \in\mathbb{R}^{6},\\
  \omega_{ij}&=\sigma\!\left(\Phi_{\mathrm{gate}}(\mathbf{m}_{ij})\right),\\
  s_{ij}&=(1-\omega_{ij})\hat{g}_{ij}+\omega_{ij}u_{ij}
  =\hat{g}_{ij}+\omega_{ij}(u_{ij}-\hat{g}_{ij}).
\end{aligned}
\label{eq:gated_fusion}
\end{equation}
Here $\Phi_{\mathrm{gate}}$ is a shared MLP, $\omega_{ij}\in(0,1)$ is the
topology interpolation weight, and $s_{ij}$ is the fused logit. The logit
$u_{ij}$ receives direct pair supervision, whereas $\rho_{ij}$ has no separate
target and only conditions the gate. Equation~\eqref{eq:gated_fusion} is a
pair-dependent convex interpolation: topology moves the geometry logit toward
$u_{ij}$ by a learned fraction of their disagreement, and has no effect when
the two streams agree. A global mixture is recovered by constraining
$\omega_{ij}=\omega$ for all pairs. The gate starts geometry-dominated, and
the resulting logits form $\mathbf{S}_{\mathrm{fuse}}=[s_{ij}]$.

Applying the sigmoid function to the fused logits yields the symmetric
pair-score matrix $\mathbf{P}=[p_{ij}]$, where
$p_{ij}=\sigma(s_{ij})=p_{ji}$ for $i\neq j$; self-pairs (diagonal entries)
are excluded throughout. Opposing-face support relations are retained
at a single operating point (Fig.~\ref{fig:fp_architecture}-d),
\begin{equation}
  \widehat{\mathcal{E}} = \{(i,j) \mid i<j,\; p_{ij} \ge \theta\},
  \label{eq:pair_decision}
\end{equation}
where $\theta$ is selected once on held-out validation data and then fixed for
test models (Sec.~\ref{sec:training}). Each pair is decided independently,
without row-capacity, mutual-best, or assignment constraints; consequently,
several retained edges may share the same face. Unlike the
multiple model-sensitive thresholds of a rule-based pipeline, $\theta$ is one
data-selected operating point on the learned pair score. The resulting support
graph is passed unchanged to the deterministic composition of
Sec.~\ref{sec:group_composition}.

\subsection{Deterministic Face-Group Composition}
\label{sec:group_composition}

Because the decisions in Eq.~\eqref{eq:pair_decision} are independent, one
face may support several others, and overlapping one-to-many relations may
jointly witness an $m$-to-$n$ wall group. Grouping should not reopen this
geometric decision with further rules. We therefore adapt the deterministic
graph composition of a prior mid-surface pipeline~\cite{ye2026midsurfer} to
the learned support set $\widehat{\mathcal{E}}$ and explicitly reject
non-bipartite components (Fig.~\ref{fig:fp_architecture}-e).

\begin{algorithm}[t]
\caption{Deterministic face-group composition}
\label{alg:group_composition}
\small
\KwIn{Predicted support relations $\widehat{\mathcal{E}}$}
\KwOut{Candidate face-group partitions $\widehat{\mathcal{H}}$ and
non-bipartite components $\widehat{\mathcal{C}}_{\mathrm{bad}}$}
$\mathsf{Adj}\leftarrow$ the undirected adjacency induced by
$\widehat{\mathcal{E}}$\;
\nllabel{line:group-graph}

Mark every face in $\mathsf{Adj}$ as uncolored and initialize
$\widehat{\mathcal{H}},\widehat{\mathcal{C}}_{\mathrm{bad}}\leftarrow
\emptyset$\;

\ForEach{uncolored face $v$ in $\mathsf{Adj}$}{
  initialize $Q\leftarrow[v]$, $\mathsf{c}(v)\leftarrow+1$,
  $\mathcal{A},\mathcal{B}\leftarrow\emptyset$, and
  $\mathsf{valid}\leftarrow\mathrm{true}$\;
  \nllabel{line:group-bfs-begin}

  \While{$Q$ is not empty}{
    pop $u$ from $Q$ and insert it into $\mathcal{A}$ or $\mathcal{B}$
    according to $\mathsf{c}(u)$\;

    \ForEach{$w\in\mathsf{Adj}[u]$}{
      if $w$ is uncolored, assign $\mathsf{c}(w)\leftarrow-\mathsf{c}(u)$
      and enqueue it; otherwise mark the component invalid when
      $\mathsf{c}(w)=\mathsf{c}(u)$\;
      \nllabel{line:group-neighbor}
    }
  }

  record $(\mathcal{A},\mathcal{B})$ in $\widehat{\mathcal{H}}$ if valid;
  otherwise record $\mathcal{A}\cup\mathcal{B}$ in
  $\widehat{\mathcal{C}}_{\mathrm{bad}}$\;
  \nllabel{line:group-output}
}
\end{algorithm}

Algorithm~\ref{alg:group_composition} constructs the support adjacency
(Line~\ref{line:group-graph}). Starting from each uncolored seed, it traverses
and two-colors the corresponding connected component while detecting
same-color conflicts
(Lines~\ref{line:group-bfs-begin}--\ref{line:group-neighbor}). It then records
each conflict-free bipartition as a candidate group and reports an invalid
component separately (Line~\ref{line:group-output}). The resulting
partitions cover the one-to-one, one-to-many, and many-to-many cases of
Sec.~\ref{sec:problem_formulation}. The algorithm neither infers missing
relations nor merges disconnected components, and it adds no learned
parameters, geometric tests, or thresholds beyond
Sec.~\ref{sec:fusion_decision}; faces without support remain unassigned.

\paragraph{\textbf{Exactness per ground-truth group}}
For a ground-truth wall group $(\mathcal{A}_k,\mathcal{B}_k)$, exact recovery
is guaranteed if the predicted relations cover and connect all its faces,
every predicted relation between its faces joins one face in $\mathcal{A}_k$
to one face in $\mathcal{B}_k$, and no relation connects the group to another
wall or a lateral face. The group is then one connected bipartite component,
whose coloring recovers $(\mathcal{A}_k,\mathcal{B}_k)$ up to exchanging the
two sides. A complete bipartite prediction is sufficient but not necessary:
connected, overlapping one-to-many patterns also suffice, whereas disjoint
one-to-one relations remain separate. If these conditions hold for every
ground-truth group and no extra lateral-only component appears, the complete group set is recovered exactly.

\paragraph{\textbf{Structural stability and scope}}
Classical face-pairing pipelines establish relations through geometric
criteria~\cite{rezayat1996midsurface} and may extend them with rule-based
completion. Once our scorer fixes $\widehat{\mathcal{E}}$, composition instead
uses only connectivity and bipartition. For a fixed support graph whose components are bipartite---each component is connected by construction, so this is precisely the \textit{connected-and-bipartite condition} named in the front matter---the unordered components are independent of traversal order, and each bipartition is unique up to exchanging its
sides---the standard property that a connected bipartite graph admits exactly
one proper two-coloring up to swapping its color classes.
In particular, the composed groups depend on the support graph only through
its components and their bipartitions; which witness edges realize them does
not affect the result. This is a
conditional structural guarantee, not error correction: a false bridge may
merge walls, a missing bridge may split one, and a side-inconsistent edge may
invalidate a component.

\begin{figure*}[t]
    \centering
    \includegraphics[width=0.9\linewidth]{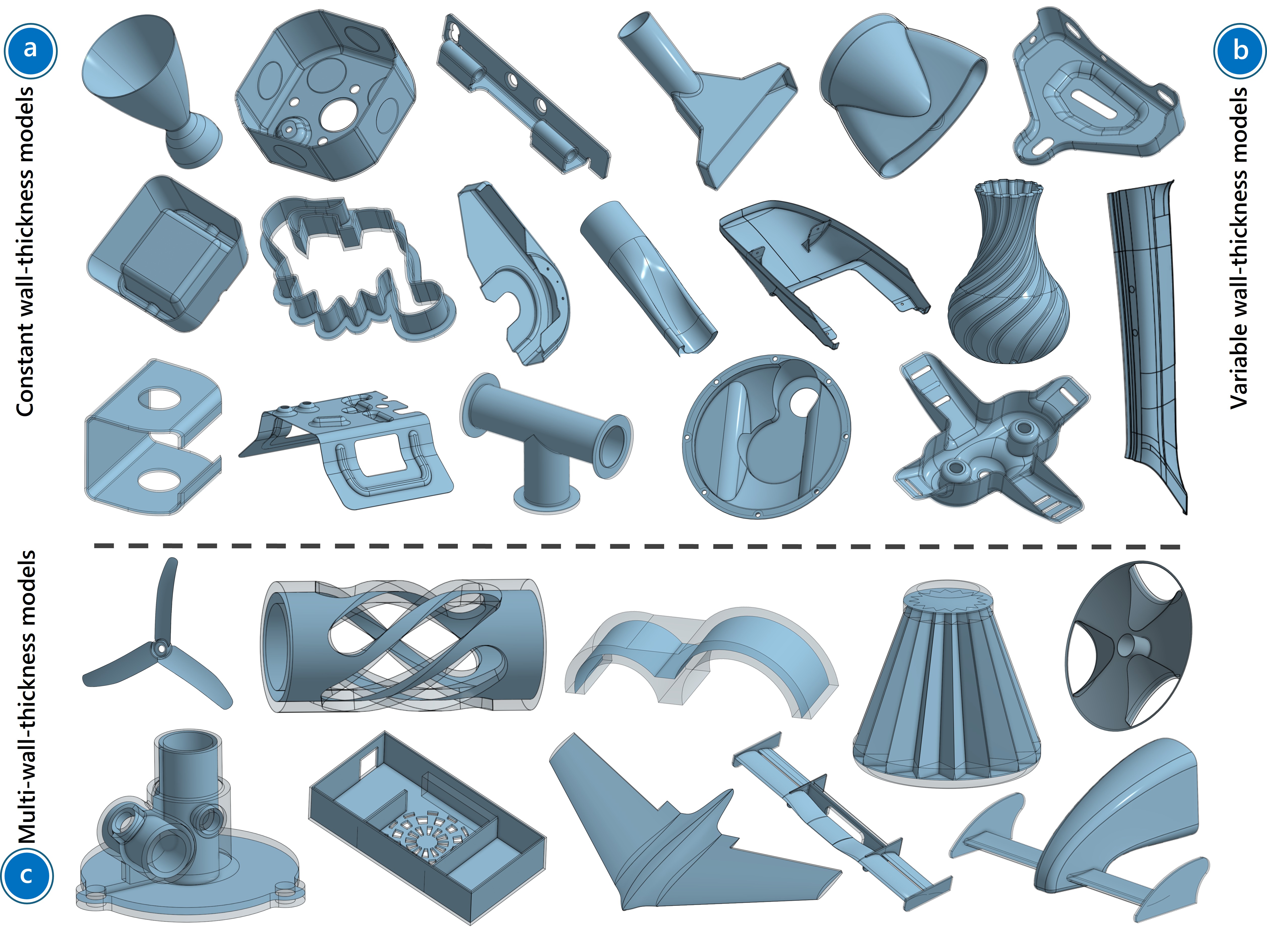}
    \vspace{-3mm}
    
    \caption{Representative models from the MidSurf dataset: (a) constant-thickness, (b) variable-thickness, and (c) multi-wall-thickness.}

    \label{fig:dataset}
    
\end{figure*}
\subsection{Training}
\label{sec:training}

\paragraph{\textbf{Loss}}
We train on all unordered face pairs: annotated pairs are positive and all
remaining pairs are negative. Let
$\mathcal{I}$ denote this pair set and $\mathcal{J}_{+}$ and
$\mathcal{J}_{-}$ the positive and negative subsets of
$\mathcal{J}\subseteq\mathcal{I}$. When both classes are present, we balance
them within each CAD model using
\begin{equation}
\begin{aligned}
  \ell_{\mathrm{bal}}(x;\mathcal{J})
  ={}&
  \frac{1}{2|\mathcal{J}_{+}|}
  \sum_{(i,j)\in\mathcal{J}_{+}}
  \operatorname{softplus}(-x_{ij})\\
  &+\frac{1}{2|\mathcal{J}_{-}|}
  \sum_{(i,j)\in\mathcal{J}_{-}}
  \operatorname{softplus}(x_{ij}).
\end{aligned}
  \label{eq:balanced_pair_loss}
\end{equation}
If only one class is present, $\ell_{\mathrm{bal}}$ reduces to the average
over that class; an empty domain has zero loss. All pairs are retained without subsampling. With
$\mathcal{I}_{\mathrm{sh}}=\{(i,j)\in\mathcal{I}\mid v_{ij}=1\}$, the total
objective is
\begin{equation}
\begin{aligned}
  \mathcal{L}_{\mathrm{total}}
  ={}&
  \ell_{\mathrm{bal}}(s;\mathcal{I})
  +0.5\,\ell_{\mathrm{bal}}(\hat{g};\mathcal{I})\\
  &+0.5\,\ell_{\mathrm{bal}}(u;\mathcal{I})
  +0.25\,\ell_{\mathrm{bal}}(a;\mathcal{I}_{\mathrm{sh}}).
\end{aligned}
  \label{eq:total_loss}
\end{equation}
Here, $s$, $\hat{g}$, $u$, and $a$ are the fused, geometry, topology, and raw
shape logits, respectively. The weights mirror the scoring hierarchy: the
deployed fused logit carries full weight, the two streams are supervised
symmetrically at half weight so that neither degenerates before fusion, and
the subordinate raw shape logit receives quarter weight. The auxiliary shape
term encourages $a_{ij}$ to align with the pair label but introduces no
standalone inference decision: at
inference, $a_{ij}$ acts only through the bounded evidence $q_{ij}$ of
Eq.~\eqref{eq:signed_shape_evidence} and the pair-conditioned correction
$r_{ij}$ of Eq.~\eqref{eq:corrected_logit}. The deterministic composition
stage has no trainable parameters and receives no group-level loss.

\paragraph{\textbf{Optimization}}
All trainable components are jointly optimized from scratch, with one CAD model
processed per update. The shape-head output layer is zero-initialized, the
conditional shape amplitudes start at $0.05$, and the gate output weights are
initialized to zero with bias $-3$, giving $\omega_{ij}\approx0.047$. We use
AdamW~\cite{loshchilov2017decoupled} with a learning rate of
$3\times10^{-4}$, betas $(0.9,0.95)$, $\epsilon=10^{-10}$, and weight decay
$10^{-4}$. The learning rate is reduced to $10^{-6}$ by cosine
annealing~\cite{loshchilov2016sgdr}, and gradients are clipped at norm $1.0$.
We train the network for 100 epochs using random seed 1003.
At each checkpoint, an exact, tie-aware sweep over the pooled validation
pairs selects the $\theta$ that maximizes F1-Score, recording the
corresponding F1-Score
and precision.
Checkpoints are ranked lexicographically by validation macro per-model
average precision (AP), then by the pooled F1-Score and precision recorded
at each checkpoint's own selected threshold; AP is threshold-free, so the
primary comparison precedes any choice of operating point; the chosen checkpoint's threshold is then fixed for test
evaluation. This selects an operating point, not a posterior calibration.

\section{MidSurf Dataset}
\label{sec:dataset}

Existing mid-surface abstraction methods are typically evaluated on small, method-specific collections without standardized annotations for variable thickness and multi-wall-thickness regions; large geometry corpora such as ABC~\cite{koch2019abc} supply raw shapes at scale but no face-pairing supervision. We therefore construct the \textit{MidSurf} dataset, comprising 1,575 CAD models with manually annotated one-to-one opposing-face support relations. The models span industrial components, academic benchmarks, and synthetically constructed corner cases.
Fig.~\ref{fig:dataset} shows representative examples from each category.

\subsection{Data Collection}
\label{sec:data_collection}

We collected CAD models from three complementary sources for diversity in geometry, topology, and industrial relevance.

\textbf{Industrial CAD Models (897 models, 57.0\%).} We collected thin-walled components from automotive, aerospace, and mechanical engineering domains via the GrabCAD online library\footnote{\url{https://grabcad.com/library}}. Models were filtered to retain those exhibiting thin-walled characteristics, with face counts ranging from 10 to 250 so that exhaustive manual pair annotation and cross-review remained tractable. This subset captures real-world geometric complexity including filleted transitions, chamfered edges, and variable-thickness regions common in industrial applications.

\textbf{Public CAD Dataset (587 models, 37.3\%).} We incorporated a randomly sampled subset of sheet-metal parts from SMCAD~\cite{ma2024adaptive}, a large-scale machining feature dataset. These models provide coverage of standard manufacturing geometries with well-defined face-pairing relationships, serving as relatively straightforward cases that establish baseline performance.

\textbf{Manually Constructed Models (91 models, 5.8\%).} Existing CAD repositories contain few controlled multi-wall-thickness examples, providing inadequate supervision for learning support relations without a single global thickness. We therefore constructed models with distinct local thicknesses.

All collected models underwent geometric validation using Parasolid's healing functions to ensure watertight B-Rep representations. Models with invalid topology or non-thin-walled characteristics were excluded; the face-count cap above bounds annotation effort and is not a capacity limit of the method. Duplicate and near-duplicate models were identified via geometric hashing and removed to prevent data leakage.

\subsection{Annotation Protocol}
\label{sec:annotation_protocol}

The MidSurf dataset provides manually annotated face-pair labels for studying the learned face-pair scorer and its downstream composition.
\label{sec:fp_annotation}

Since the STEP format does not assign explicit identifiers to B-Rep faces, our annotation tool and the network's data-loading pipeline share the same face-indexing logic provided by the \texttt{occwl} library~\cite{occwl2021}, ensuring ID consistency between face labels and training data.

We developed an interactive annotation tool built on PyVista~\cite{sullivan2019pyvista} and PyQt5. After loading a STEP model, the annotator selects faces and records each opposing-face support relation as a two-element tuple of face IDs, written line by line to a plain-text file; the two IDs correspond to opposite faces on either side of the wall.

The manual annotation of all 1,575 models was systematically divided among three annotators. A separate color-coded verification tool was developed for quality control: it reads the annotation file, assigns a unique color to each pair, and renders the highlighted faces on the 3D model, enabling rapid visual inspection of annotation correctness. 
All annotators participated in multiple rounds of cross-review and correction before models were admitted to the final dataset. To assess consistency of the pair annotations, 50 randomly selected models were independently labeled by two annotators, achieving an inter-annotator agreement (Cohen's Kappa) of 0.91. On average, annotating a single model required approximately 3--5 minutes depending on geometric complexity.

\begin{table}[t]
    \setlength{\tabcolsep}{4pt}
    \centering
    \footnotesize
    \caption{Overview of the MidSurf dataset. The dataset comprises 1,575
CAD models with stratified train/val/test splits.}
    \label{tab:dataset_statistics}
    \begin{tabular*}{\columnwidth}{@{\extracolsep{\fill}}lc@{}}
        \toprule
        \textbf{Metric} & \textbf{Value} \\
        \midrule
        Total models & 1,575 \\
        Total faces & 207,187 \\
        Total annotated support relations & 28,514 \\
        \midrule
        Faces per model (min / max / mean) & 3 / 250 / 131.5 \\
        Support annotations per model (min / max / mean) & 1 / 91 / 18.1 \\
        \midrule
        Constant-thickness models & 415 (26.3\%) \\
        Variable-thickness models & 1,160 (73.7\%) \\
        \addlinespace
        Multi-wall-thickness models & 142 (9.0\%) \\
        \multicolumn{2}{@{}l}{\quad\scriptsize\itshape subset of the variable-thickness category} \\
        \midrule
        Train / Val / Test split & 1,102 / 237 / 236 \\
        \bottomrule
    \end{tabular*}
\end{table}
\subsection{Dataset Statistics}
\label{sec:dataset_statistics}

Table~\ref{tab:dataset_statistics} summarizes the MidSurf dataset statistics. The dataset exhibits broad coverage across model complexities, with face counts ranging from 3 to 250 (mean: 131.5) and pair annotations per model ranging from 1 to 91 (mean: 18.1).

The dataset is partitioned into training (70\%), validation (15\%), and test (15\%) sets using stratified sampling to balance constant/variable-thickness and multi-wall-thickness models across splits. Cross-split leakage from related models is prevented at collection time, where geometric hashing removes duplicates and near-duplicates (Sec.~\ref{sec:data_collection}).

\textbf{Availability.} The MidSurf dataset, including all CAD models in STEP format and face-pairing annotations, will be publicly released upon paper acceptance. We will also release the interactive face-pair annotation tool and train/val/test split specifications to facilitate reproducibility and future research in learning-based mid-surface abstraction.

\vspace{-2mm}
\section{Experiments}
\label{sec:experiments}

We evaluate \toolName{} through comprehensive experiments on the MidSurf dataset (Sec.~\ref{sec:dataset}). We first describe the implementation setup and the evaluation metrics, then present the quantitative analysis, the comparison with existing methods, ablation studies validating our design choices, and a case study applying the generated mid-surfaces to finite element analysis.

\textbf{Implementation.} We implement \toolName{} in Python 3.10 using
PyTorch 2.11 and PyTorch Geometric 2.8 for the neural components. B-Rep
geometry processing is handled by the \texttt{occwl} library~\cite{occwl2021}
built on Open CASCADE. All experiments are conducted on a 64-bit Windows 11
workstation equipped with a 24-core Intel Core Ultra 7 270K Plus CPU, 64~GB
RAM, and a single NVIDIA GeForce RTX 5090 GPU with 32~GB memory. GPU training
uses the CUDA 12.8 build of PyTorch. Training details are provided in Sec.~\ref{sec:training}.

\vspace{-1mm}
\subsection{Evaluation Metrics}
\label{sec:metrics}

The learned face-pair scorer and the deterministic composition stage form the
face-pairing pipeline but solve two different parts of the task. We therefore
report their performance at two levels: scorer metrics measure the predicted
support relations, whereas the full-pipeline metric measures the mid-surface
that \toolName{} finally delivers.

\textbf{Learned Face-Pair Scorer Metrics.}
\begin{itemize}
    \item \textit{Precision}: the proportion of predicted support relations that are annotated as positive.
    \item \textit{Recall}: the proportion of annotated support relations recovered by the scorer.
    \item \textit{F1-Score}: the harmonic mean of precision and recall.
\end{itemize}
These metrics are computed over unordered face pairs immediately after the
decision in Eq.~\eqref{eq:pair_decision}, before deterministic composition, and
therefore isolate the learned scorer. Counts are pooled over pairs across the
test set (micro-averaging), so each model contributes in
proportion to the number of annotated relations it contains. Since the annotations record one specific set of
one-to-one relations per wall, a predicted relation whose two faces bound the
same wall from opposite sides, but which was not itself annotated, still
counts as a false
positive; the reported precision is therefore conservative.

\textbf{Full-Pipeline Metric.}
Following MidSurfer~\cite{ye2026midsurfer}, we use \textit{Completion Rate} to evaluate
the mid-surface produced by the complete pipeline. It is assessed manually: an
expert checks whether the mid-surface of a model shows topological omissions or
additions with respect to the input solid, and the model counts as completed
only when neither occurs. The Completion Rate of a set of models is the
proportion of its models that are completed. All judged mid-surfaces are
constructed by the same downstream API (Sec.~\ref{sec:method_overview}) from
each method's own face groups, so only the constructor is shared across
methods.

\vspace{-1mm}
\subsection{Quantitative Analysis}
\label{sec:quantitative}

\begin{figure}[t]
    \centering
    \includegraphics[width=\linewidth]{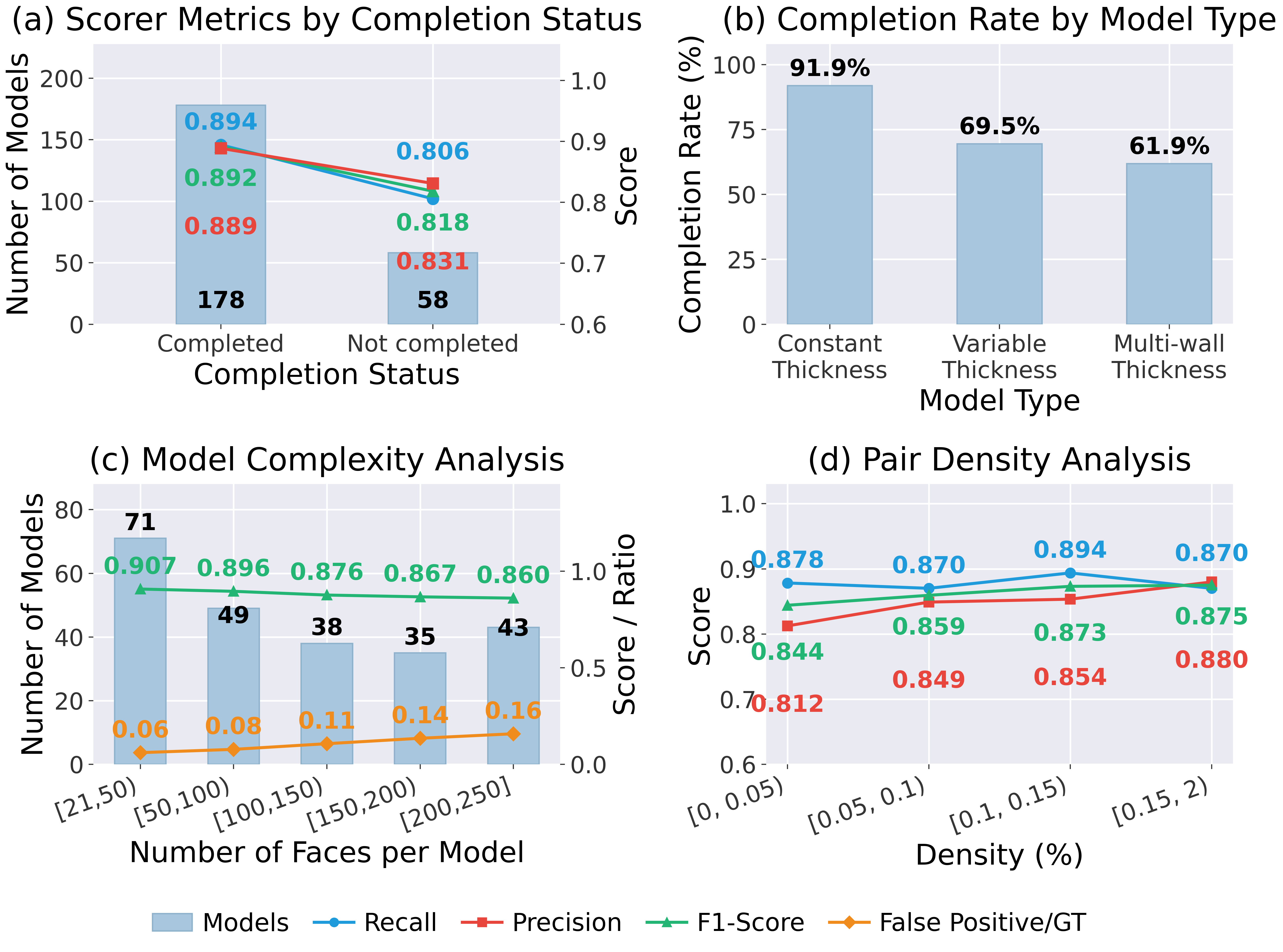}
    \vspace{-8mm}
    \caption{Quantitative evaluation at the scorer and full-pipeline levels.
    (a) Full-pipeline completion status, together with scorer recall,
    precision, and F1-Score on the completed and the not-completed models.
    (b) Completion Rate per model type (multi-wall is a subset of variable-thickness). (c) Scorer F1-Score and
    false-positive-to-ground-truth ratio versus model complexity. (d) Scorer recall, precision,
    and F1-Score across pair-density intervals.}
    \vspace{-5mm}
    \label{fig:face_pairing_quantitative}
\end{figure}

We analyze the two stages along four dimensions. Completion Rate and model type
characterize the complete pipeline, whereas model complexity and pair
density isolate the learned face-pair scorer. Face count and pair density
are inversely coupled by construction, so panels (c) and (d) offer two views
of one axis rather than two independent factors; all subset scorer metrics in this
section are pooled over pairs, as in Sec.~\ref{sec:metrics}, whereas
Completion Rate is pooled over models. The completion analysis additionally
relates the full-pipeline outcome to the scorer metrics that produce it.

\textbf{Completion Rate Analysis.}
Fig.~\ref{fig:face_pairing_quantitative}-a first evaluates the complete
pipeline by splitting the test set according to the completion judgment of its
mid-surface. Among 236 test models, 178 (75.42\%) are completed, that is, their
mid-surfaces show neither topological omissions nor additions.
We then inspect scorer performance within the same two groups; the split is
descriptive, since completion is judged downstream of the scorer. The completed
models obtain an F1-Score of 0.892 against 0.818 for the rest and lead on
both axes: recall 0.894 against 0.806 and precision 0.889 against 0.831.
This is consistent with the manual criterion, which penalizes omissions
and additions alike---a mid-surface fails for either. Recall stays below one even on
completed models, consistent with Sec.~\ref{sec:group_composition}: a group is
recovered once the retained relations form a connected bipartite witness of it,
so not every annotated relation has to be predicted.

\textbf{Model Type Analysis.}
Fig.~\ref{fig:face_pairing_quantitative}-b reports the full-pipeline
Completion Rate per model type. It declines with thickness complexity:
91.94\% (57/62) for constant-thickness models, 69.54\% (121/174) for
variable-thickness ones, and 61.90\% (13/21) for the multi-wall subset of the
variable-thickness category. Wall thickness is not explicitly supplied to
the scorer; the relations must be inferred from geometry and topology before
deterministic composition organizes them into face groups. The lower rates on
the harder categories therefore reflect the greater difficulty of the complete
task rather than dependence on a prescribed thickness value.

\textbf{Model Complexity Analysis.}
We next isolate the learned face-pair scorer and employ the number of faces as a
proxy for geometric complexity (Fig.~\ref{fig:face_pairing_quantitative}-c).
As the face count increases from $[21, 50)$ to $[200, 250]$, the F1-Score
gradually decreases from 0.907 to 0.860, while the
false-positive-to-ground-truth ratio (False Positive/GT) rises steadily from
0.06 to 0.16.
Both curves describe the same effect: as the face count grows without a
proportional increase in pair count, the quadratic expansion of the candidate
space ($O(N^2)$) makes the scorer more
susceptible to false-positive predictions, although the ratio grows far more
slowly than the candidate space itself, so the scorer suppresses most of the
added negatives.

\textbf{Pair Density Analysis.}
Finally, we examine the learned scorer with respect to pair density,
$100\,|\mathcal{E}^*| / \binom{N}{2}$ (in \%), where $N$ denotes the number of
faces and $\mathcal{E}^*$ is the set of annotated support relations. This metric characterizes
the sparsity of supervised positives in the quadratic pair domain. As shown in
Fig.~\ref{fig:face_pairing_quantitative}-d, the lowest-density range yields a
precision of 0.812 and a recall of 0.878. At higher densities, precision rises steadily to
0.880 while recall varies without trend within 0.870--0.894: sparsity costs precision rather
than coverage, so false positives remain the bottleneck in the sparsest
regime.

\begin{figure*}[t]
    \centering
    \includegraphics[width=0.8\linewidth]{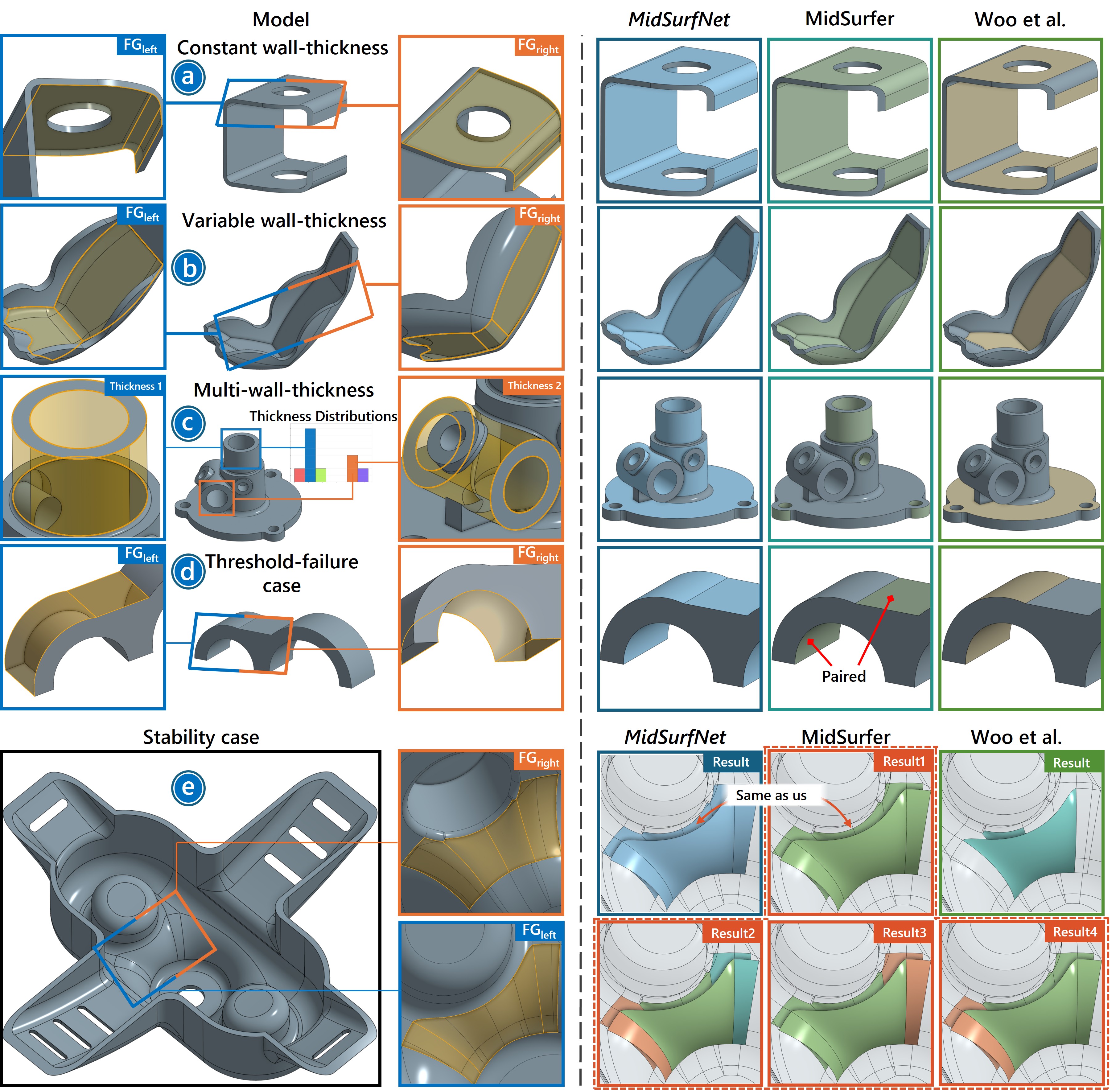}
    \vspace{-2mm}
    \caption{Face-pairing comparison. (a)--(c) One model per
    category---constant, variable, and multi-wall thickness---with the
    reference face groups (FG$_{\mathrm{left}}$/FG$_{\mathrm{right}}$),
    composed from the one-to-one annotations, and each method's pairing
    result; the inset in (c) shows the thickness distribution, one peak per
    local wall thickness; (d) a threshold-failure case in which
    relaxed rule conditions pair non-opposing faces; (e) a grouping-stability
    case in which the rule-based pipelines admit several distinct groupings
    of one region (Results 1--4, only Result~1 matching ours), while the
    deterministic composition returns a single partition.}
    \vspace{-5mm}
    \label{fig:QS_FP}
\end{figure*}
\subsection{Comparative Experiments}
\label{sec:comparison}
\label{sec:fp_comparison}

We compare \toolName{} with two families of alternatives. Rule-based systems
are complete pipelines, so Sec.~\ref{sec:comparison_rule} evaluates them at
both levels---scorer metrics and Completion Rate. The learned baselines of
Sec.~\ref{sec:comparison_learned} replace only the face-pair scorer, while
the deterministic composition and the downstream constructor stay fixed; they
are therefore compared at the pair level, where the architectures actually
differ.

\subsubsection{Comparison with Rule-Based Methods}
\label{sec:comparison_rule}

We compare \toolName{} with two representative rule-based baselines at both the face-pair-scorer and full-pipeline levels:
\begin{itemize}
    \item \textbf{Woo et al.}~\cite{woo2014divide}: A divide-and-conquer approach that recursively partitions the model and applies local pairing rules.
    \item \textbf{MidSurfer}~\cite{ye2026midsurfer}: The strongest evaluated rule-based method, with automatic thickness detection and hierarchical filtering.
\end{itemize}

\begin{table}[t]
    \centering
\setlength{\tabcolsep}{3pt}
    \footnotesize
    \caption{Comparison at the face-pair-scorer and full-pipeline levels.
    Precision, recall, and F1-Score are computed over unordered face pairs
    before deterministic composition; Completion Rate is assessed on the
    final mid-surfaces. A dash (--) denotes an unsupported model category;
    its models count as not completed in the enclosing rows. Multi-wall is
    a subset of the variable-thickness category, so the category rows are
    not additive.}
    \label{tab:fp_completion_transposed}
    \begin{tabular*}{\columnwidth}{@{\extracolsep{\fill}}lccc@{}}
        \toprule
        \textbf{Metric / subset} & Woo et al. & MidSurfer & \textbf{\toolName{}} \\
        \midrule
        \multicolumn{4}{@{}l}{\textit{Pair-level metrics}} \\
        \quad Precision            & 0.1294 & 0.5576 & \textbf{0.8745} \\
        \quad Recall               & 0.1418 & 0.7537 & \textbf{0.8719} \\
        \quad F1-Score             & 0.1353 & 0.6410 & \textbf{0.8732} \\
        \midrule
        \multicolumn{4}{@{}l}{\textit{Full pipeline: Completion Rate (\%)}} \\
        \quad Constant-thickness         & 24.19 & 80.65 & \textbf{91.94} \\
        \quad Variable-thickness         & -- & 43.68 & \textbf{69.54} \\
        \quad Multi-wall-thickness       & --  & --  & \textbf{61.90} \\
        \addlinespace
        \quad All models                 & 6.36 & 53.39 & \textbf{75.42} \\
        \bottomrule
    \end{tabular*}
\end{table}

\textbf{Pair-Level Comparison.}
Table~\ref{tab:fp_completion_transposed} first compares the unordered pair
predictions before deterministic composition. Woo et al.'s method obtains 0.1294 precision, 0.1418 recall, and an F1-Score of 0.1353. MidSurfer raises recall to 0.7537 while precision stays at 0.5576---a profile its two-stage rule
design explains: after the initial distance--normal--overlap filtering,
unmatched faces are revisited through adjacency-guided complementary pairing
under relaxed conditions~\cite{ye2026midsurfer}, which recovers difficult
positives but also retains geometrically plausible non-pairs. \toolName{}
reaches a balanced precision of 0.8745 and recall of 0.8719 (F1-Score
0.8732, 23.22 points above MidSurfer): the geometry stream evaluates direct geometric
compatibility, the attributed-topology stream supplies structural context,
and the pair-conditioned gate adjusts their weights for each pair, rejecting
locally plausible but structurally inconsistent negatives without sacrificing
true pairs when either evidence source turns ambiguous.

We then evaluate whether each method's pair decisions ultimately yield a
complete mid-surface. Overall, the three methods complete 6.36\%,
53.39\%, and 75.42\% of the test models; the category rows of
Table~\ref{tab:fp_completion_transposed} locate the difference, and
Fig.~\ref{fig:QS_FP} accompanies them: panels (a)--(c) show one model per
category, panel (d) a threshold-failure case, and panel (e) a
grouping-stability case we return to below.

\textbf{Constant-Thickness Models.}
Woo et al.'s method completes 24.19\% of this category, MidSurfer
80.65\%, and \toolName{} 91.94\% (Fig.~\ref{fig:QS_FP}-a). Despite the
constant wall thickness these models remain geometrically complex, so the
comparison tests more than the nominal thickness assumption; annotation
ambiguity also accounts for part of the residual gap
(Sec.~\ref{sec:discussion}).

\textbf{Variable-Thickness Models.}
Woo et al.'s constant-thickness design does not carry over: its
maximal-volume decomposition can produce abnormal half-space partitions,
exploding the number of half-spaces or crashing outright
(Fig.~\ref{fig:QS_FP}-b). MidSurfer improves this category to
43.68\% by incorporating local topological information, but when
variable thickness drives face normals away from the projection directions,
its thresholds again become the bottleneck (Fig.~\ref{fig:QS_FP}-d).
\toolName{} completes 69.54\%: the topological context fed to the network
lets it learn support relations across continuous geometric variation rather
than through fixed-threshold decisions.

\textbf{Multi-Wall-Thickness Models.}
A single part may contain several local wall thicknesses, each forming a
separate peak in the wall-thickness histogram (Fig.~\ref{fig:QS_FP}-c). The
evaluated implementations of Woo et al.\ and MidSurfer cannot handle this
category, whereas \toolName{} completes 61.90\%. Rather than hard-filtering
pairs through an estimated thickness, the two streams evaluate each pair's
opposing likelihood directly, so no single global threshold has to reconcile
multiple local thicknesses.

\textbf{Grouping Stability.}
None of the metrics above can express one further difference: whether a
method maps one part to one result. Because rule-based grouping depends on
thickness-mode selection and merge order, the same region admits a set of
plausible groupings---Fig.~\ref{fig:QS_FP}-e shows several (Results 1--4)
from the rule-based pipelines, of which only Result~1 coincides with ours.
This is the comparison that concerns the \textit{deterministic face-group
composition} itself: for a fixed support graph it returns a single partition,
independent of processing order and unique up to exchanging the side labels
(Sec.~\ref{sec:group_composition}). Sec.~\ref{sec:fea_case} evaluates what this spread costs downstream, where alternative compositions of the same parts diverge in simulation fidelity.

\subsubsection{Comparison with Learning-Based Baselines}
\label{sec:comparison_learned}

No prior learning-based face-pairing method exists for direct comparison, so
we construct three baselines that replace the learned face-pair scorer while
keeping everything else fixed. Each targets one design question:
\begin{itemize}
    \item \textbf{XGBoost}~\cite{chen2016xgboost}: Gradient-boosted trees on the concatenated handcrafted pair evidence---the criterion vector $\mathbf{c}_{ij}$, the shape vector $\mathbf{x}_{ij}$, and the structural statistics $\boldsymbol{\eta}_{ij}$---testing whether a strong tabular learner on the same inputs suffices without deep learning.
    \item \textbf{MLP}: The same concatenated features fed to a plain multilayer perceptron trained with the balanced pair loss of Eq.~\eqref{eq:balanced_pair_loss}, testing whether a flat network over the same handcrafted pair evidence matches the structured scorer.
    \item \textbf{BRT + pair head}~\cite{zou2025bringing}: BRT face embeddings trained end to end on our data and scored by a symmetric head over $[\,\mathbf{e}_i{+}\mathbf{e}_j \,\|\, |\mathbf{e}_i{-}\mathbf{e}_j| \,\|\, \mathbf{e}_i{\odot}\mathbf{e}_j\,]$, testing whether a general-purpose B-Rep encoder replaces the criteria-grounded design.
\end{itemize}

All baselines use the identical data split, the complete unordered-pair
domain, the same class weighting, and an operating threshold selected
independently on the validation set; hyperparameters are tuned on validation
under a comparable budget. Because these baselines replace only the scorer,
Table~\ref{tab:learned_baselines} reports pair-level metrics; the shared
composition stage and constructor are evaluated in
Sec.~\ref{sec:comparison_rule} and Sec.~\ref{sec:fea_case}.
\begin{table}[t]
\centering
\setlength{\tabcolsep}{3pt}
\footnotesize
\caption{Learning-based baselines. Each replaces the learned face-pair
scorer while keeping the pair domain, data split, class weighting, and
threshold protocol fixed; metrics are pair-level, as the composition stage is
common to all (Sec.~\ref{sec:comparison_learned}).}
\label{tab:learned_baselines}
\begin{tabular*}{\columnwidth}{@{\extracolsep{\fill}}l c c c@{}}
\toprule
\textbf{Baseline} & \textbf{Precision} & \textbf{Recall} & \textbf{F1-Score} \\
\midrule
XGBoost (handcrafted features) & 0.8542 & 0.7382 & 0.7920 \\
MLP (handcrafted features) & 0.7987 & 0.8139 & 0.8062 \\
BRT + pair head & 0.3449 & 0.5940 & 0.4364 \\
\midrule
\textbf{Full Model (Ours)} & \textbf{0.8745} & \textbf{0.8719} & \textbf{0.8732} \\
\bottomrule
\end{tabular*}
\vspace{-2mm}
\end{table}

XGBoost and the MLP achieve F1-Scores of 0.7920 and 0.8062,
respectively, using the same fixed pair descriptors but without the proposed
dual-stream architecture. The fact that two different learners both achieve
an F1-Score around 0.8 indicates that the annotated data provide a consistent
learning signal and that the criterion, shape, and structural pair evidence
is already discriminative. XGBoost behaves conservatively: the criterion features let it identify
pairs with strong geometric evidence, but it recovers few relations among
small faces with ambiguous geometry. The MLP, by contrast, can learn a smoother joint decision boundary over the associated continuous descriptors, thereby recovering more positive examples at the cost of a certain precision loss.

BRT with a symmetric pair head obtains only 0.3449 precision, 0.5940 recall, and an F1-Score of 0.4364. Its substantially higher recall than precision suggests
that general-purpose face embeddings can retrieve some plausible pairs but
cannot reliably reject the many non-pairs in the complete unordered-pair
domain. This task requires more than describing the two faces independently:
it also requires their relative physical compatibility and their structural
roles in the solid. Our full model explicitly represents these factors through
the criterion-baseline geometry stream, conditional shape correction,
attributed-topology context, and pair-conditioned fusion. It consequently
achieves
balanced precision and recall of 0.8745 and 0.8719, yielding an F1-Score of
0.8732---6.7 points above the strongest learned baseline. Since BRT and our method differ
in both representation and architecture, the contribution of each individual
component is examined separately through the controlled ablations in
Sec.~\ref{sec:ablation}.

\begin{table}[t]
\centering
\setlength{\tabcolsep}{3pt}
\footnotesize
\caption{Ablation of the learned face-pair scorer. Each row removes or
replaces the named mechanism from the full model; \textit{Geometry Stream
Only} and \textit{Topology Stream Only} keep a single stream, and
\textit{Global Fusion} replaces only the Pair-Conditioned Gate. All variants evaluate the same complete unordered-pair
domain and use an independently validation-selected operating threshold.}
\label{tab:ablation_FP}
\begin{tabular*}{\columnwidth}{@{\extracolsep{\fill}}l c c c@{}}
\toprule
\textbf{Variant} & \textbf{Precision} & \textbf{Recall} & \textbf{F1-Score} \\
\midrule
\multicolumn{4}{@{}l}{\textit{Geometry}} \\
\quad W/O Criterion & 0.7984 & 0.8408 & 0.8191 \\
\quad W/O Shape & 0.8337 & 0.8616 & 0.8474 \\
\addlinespace
\quad Geometry Stream Only & 0.8511 & 0.7789 & 0.8134 \\
\midrule
\multicolumn{4}{@{}l}{\textit{Topology}} \\
\quad W/O Topology GNN & 0.9081 & 0.7562 & 0.8252 \\
\addlinespace
\quad W/O Endpoint Relation & 0.8054 & 0.8700 & 0.8365 \\
\quad W/O Shortest-Route & 0.8163 & 0.8790 & 0.8465 \\
\quad W/O Structural Statistics & 0.8481 & 0.8604 & 0.8542 \\
\addlinespace
\quad W/O $\mathbf{R}_{\mathrm{topo}}$ & 0.8009 & 0.8556 & 0.8273 \\
\quad Topology Stream Only  & 0.7531 & 0.8394 & 0.7939 \\
\midrule
\multicolumn{4}{@{}l}{\textit{Fusion}} \\
\quad Global Fusion ($\omega_{ij}{=}\omega$) & 0.8702 & 0.8538 & 0.8619 \\
\quad \textbf{Full Model (Ours)}
  & \textbf{0.8745} & \textbf{0.8719} & \textbf{0.8732} \\
\bottomrule
\end{tabular*}
\vspace{-2mm}
\end{table}

\begin{figure*}[t]
    \centering
    \includegraphics[width=\textwidth]{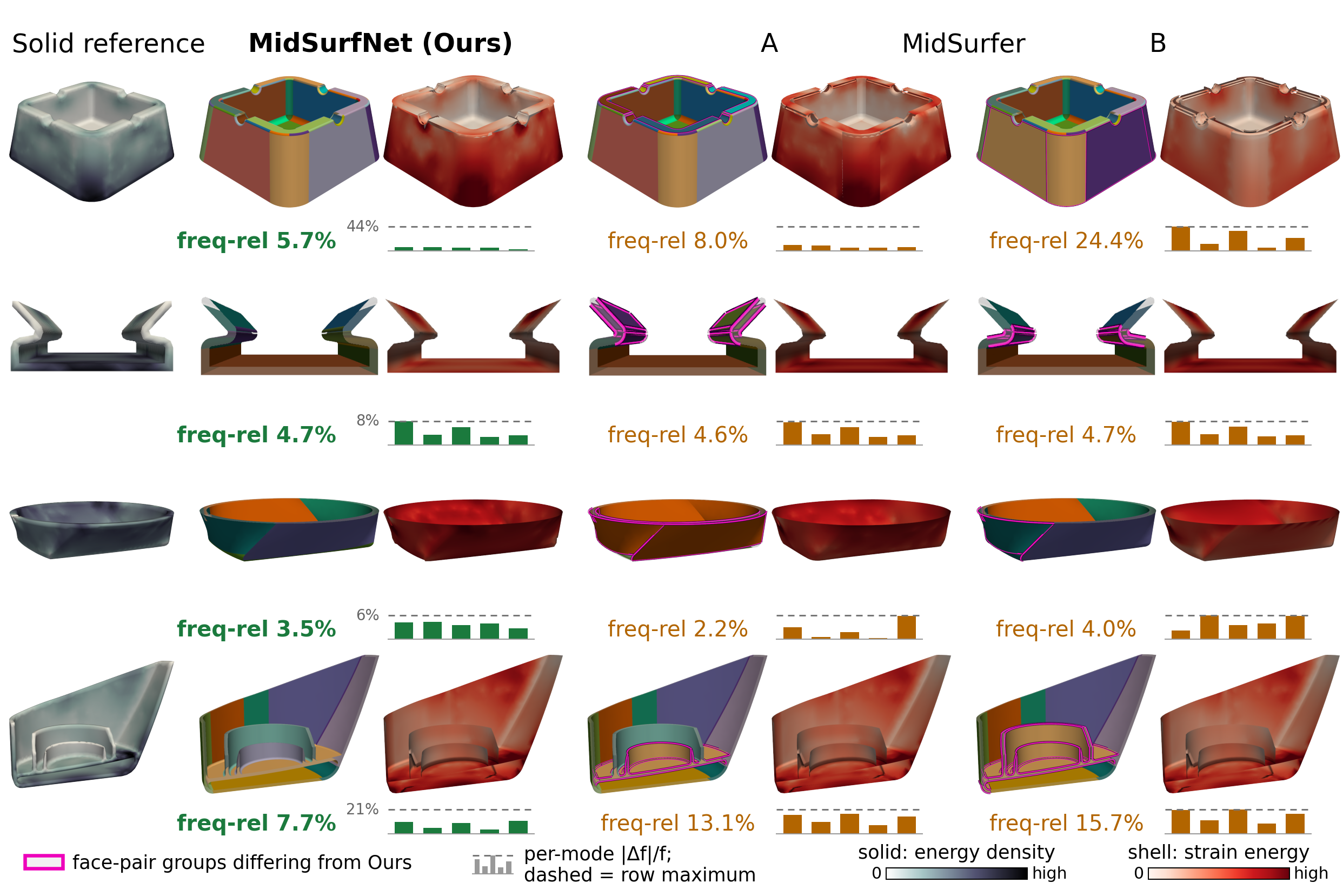}
    \vspace{-8mm}
    \caption{Free--free modal validation of final mid-surfaces. Each row is one model: the solid reference, then the face-pairing result and modal strain-energy heatmap for \toolName{} and for two alternative MidSurfer compositions (A, B) of the same part---two samples from an outcome set that is not limited to two; magenta outlines mark face-pair groups that differ from ours. Below each group, freq-rel is the mean relative error of the first five flexible free--free modes against the solid reference (lower is better), the bar charts break this error down per mode on a row-shared axis (shorter is better; the dashed line marks the row-wide maximum), and the shell heatmaps of a row share one color scale (darker red = higher modal strain energy); the solid reference uses its own white-to-black energy-density scale.}
    \label{fig:fea_case}
\end{figure*}
\subsection{Ablation Study}
\label{sec:ablation}
\label{sec:feature_ablation}

We organize the ablation around the three design questions of
Secs.~\ref{sec:neural_face_pairing} and~\ref{sec:fusion_decision}, rather than
removing individual descriptor coordinates or network layers: whether the
classical pair criteria and the shape evidence divide the geometric work as
intended, through which mechanisms the attributed-topology stream contributes,
and whether geometry--topology arbitration must be pair-conditioned. Each
variant removes or replaces only the named mechanism of the full model,
keeping the complete unordered-pair domain, data split, training budget, and
checkpoint-selection rule, and re-selecting its operating threshold on
validation (Table~\ref{tab:ablation_FP}).

\textbf{Geometry evidence.}
Removing the continuous criterion features causes the largest single-mechanism
drop (F1-Score $0.8732\!\rightarrow\!0.8191$): without distance, normal, and
overlap evidence, faces of similar shape or similar structural role are
readily misidentified as pairs, and precision falls markedly. Even within a
learning-based formulation, the classical criteria thus remain the least
replaceable evidence for face pairing. Removing the conditional shape
correction lowers precision by 4.1 points at nearly unchanged recall (0.862).
Shape evidence plays an auxiliary role in the geometry stream: it rejects
candidates whose trimmed shapes cannot bound the same wall, and it recovers
borderline small elongated faces for which the distance, normal, and overlap criteria cannot hold at once (Eq.~\eqref{eq:conditional_shape_weights}).

\textbf{Topology mechanisms.}
The topology ablations separate a substrate from its readouts. Removing the
\textit{Topology GNN} costs 4.8 points, keeping precision high while recall
collapses (0.908/0.756): without message passing, faces with strong geometric
cues are still matched, but in regions with many faces and complex topology
their neighbors score systematically lower and are hard to recall. Removing
the \textit{Endpoint Relation}, the \textit{Shortest-Route} context, or the
\textit{Structural Statistics} instead leaves recall largely unchanged while
precision drops by 2.6 to 6.9 points, which matches their shared function: all
three act as filters in topological reasoning, eliminating candidates that
are topologically similar yet do not belong to the same wall---in particular separating lateral faces from wall faces where the distance criterion alone fails. The
most instructive row is \textit{W/O} $\mathbf{R}_{\mathrm{topo}}$ ($-4.6$,
with precision $-7.4$): the conditioning signal never scores a pair itself,
yet removing it hurts more than removing any single readout. It informs the
gate's trust in the topology verdict; without it, arbitration is blind on
exactly the pairs topology should veto. What it contributes is
reliability information, not additional evidence.

\textbf{Stream complementarity and fusion.}
Removing an entire stream isolates the other branch. On its own, the topology
branch trades precision for recall (0.8394 recall at 0.7531 precision;
\textit{Topology Stream Only}), whereas the geometry branch shows the opposite
profile (0.7789 recall at 0.8511 precision; \textit{Geometry Stream Only}):
geometric evidence acts as a hard filter that validates candidate pairs,
while topological evidence acts as a soft constraint that recovers candidates
geometry overlooks. The full model dominates both single streams on both axes simultaneously
(precision 0.8745 against 0.8511, recall 0.8719 against 0.8394), so the gate
is not trading one axis for the other. Constraining the gate to a global mixture
(\textit{Global Fusion}, $\omega_{ij}{=}\omega$) forfeits 1.1 points, almost
entirely recall ($-1.8$ at $-0.4$ precision): a single mixing ratio is a
compromise across regimes, whereas the pair-conditioned gate reallocates
trust exactly on the borderline pairs whose evidence balance is atypical.

\subsection{Case Study: FEA on Generated Mid-surfaces}
\label{sec:fea_case}

Face pairing is a means to an end: the composed face groups determine the mid-surface model that downstream simulation consumes. As reported in Sec.~\ref{sec:comparison}, rule-based pairing fails to deliver a topologically complete mid-surface on a large fraction of the test set, and the resulting shell model inherits the missing or spurious walls. This case study therefore examines the subtler question that remains once completion succeeds: how much does the composition itself affect simulation fidelity?

We select four representative thin-walled models from the MidSurf test set. For each model, the thresholded support relations are composed into a candidate variable-cardinality $m$-to-$n$ partition (Sec.~\ref{sec:group_composition}) and passed to the mid-surface API of Siemens NX, which constructs the mid-surface patches group by group; each patch carries the wall thickness implied by its pair distances, and the assembled shell model is meshed and solved. We deliberately probe with free--free modal analysis: it requires no boundary-condition choices, so the computed natural frequencies reflect only the stiffness and mass distribution of the shell model itself. The reference is a solid finite-element model of the original part solved under the same protocol, and we report \emph{freq-rel}, the mean relative error of the first five flexible natural frequencies (rigid-body modes excluded) against this reference; the bar charts in Fig.~\ref{fig:fea_case} break the error down mode by mode on a row-shared axis. Because rule-based pairing depends on globally selected thickness modes and merge heuristics, the same part does not map to a unique composition but to a set of plausible outcomes---a set that may well include the composition our method produces. Columns A and B display, for each model, two \emph{other} members of this outcome set; the set is not limited to two, and which member the heuristic returns is not predictable in advance. Magenta outlines mark the face-pair groups where each displayed outcome differs from ours.

Fig.~\ref{fig:fea_case} is read as follows. Freq-rel measures deviation from the solid reference, so \emph{lower is better}: a value of zero would mean the shell model reproduces the reference frequencies exactly. The bar charts follow the same direction---shorter bars mean smaller per-mode error, and since the three charts of a row share one axis (the dashed line marks the row-wide maximum), bar heights compare directly across compositions. The heatmaps visualize where modal strain energy concentrates: darker red indicates higher strain energy on the shells, and darker shading indicates higher energy density on the solid reference. The three shell heatmaps of a row are likewise normalized to one shared scale, so the shell panels of a row are directly
comparable; a faithful shell model should reproduce the reference's energy pattern.

Under this reading, the deterministic composition of \toolName{} yields a single shell model per part, and its freq-rel stays within 3.5--7.7\% across all four models. The alternative compositions span 2.2--24.4\%: on the second row the three compositions are practically indistinguishable, and on the third row composition A is even closer to the reference than ours (2.2\% against 3.5\%); but on the first row composition B more than quadruples the error to 24.4\%, and on the fourth row the alternatives raise it from 7.7\% to 13.1\% and 15.7\%. The per-mode bars show that when degradation occurs, it is not confined to a single mode but spreads across the first five flexible modes. Crucially, nothing in the shell models themselves reveals which case one is in: every composition produces a complete, visually plausible mid-surface, and its fidelity is exposed only by the solid reference solve---precisely the computation that shell modeling is meant to avoid.

For engineering practice this asymmetry matters more than the average error. A deterministic composition gives one reproducible model that can be validated consistently across design iterations, whereas a heuristic composition may return any member of a set of outcomes whose spread---already 2.2\% to 24.4\% among the sampled alternatives---is not known in advance. This selected-case analysis illustrates the downstream sensitivity to group composition once a candidate partition is supplied; it is not an exact-group benchmark.

\section{Discussion and Limitations}
\label{sec:discussion}

\toolName{} advances mid-surface abstraction, yet its present form has clear
boundaries. We discuss them at three levels---what the pipeline can promise,
which inputs it handles poorly, and what the evaluation can and cannot yet
measure.

\subsection{The Composition Contract and the Asymmetry of Errors}
\label{sec:limit_error_propagation}

The determinism of the face-group composition is a conditional guarantee: it
organizes whatever support graph the scorer delivers, and its
order-independence and uniqueness hold only while each component remains
connected and bipartite (Sec.~\ref{sec:group_composition}). This contract
makes the scorer's two error types sharply asymmetric. A missed relation is
often harmless: a group is still recovered exactly while the retained
relations form a connected bipartite witness of it, which is why completed
models tolerate recall below one (Sec.~\ref{sec:quantitative}). A false
relation, in contrast, can bridge two walls and corrupt entire groups by
merging them or breaking bipartiteness. The ablation mirrors this asymmetry: the topology readouts act chiefly as
filters
(Sec.~\ref{sec:ablation}).

The deeper cause is architectural: the scorer decides every pair
independently, while correctness is a property of the whole graph, so
violations surface only downstream as composition failures. The predicted support evidence already permits interactive inspection
before mid-surface generation; staged rule-based pipelines share this
dependency on upstream decisions~\cite{woo2014divide,ye2026midsurfer}. Two
learned remedies appear feasible. First, weighting false positives above
omissions in the pairwise loss mirrors the cost asymmetry. Second, letting
training see graph-level consequences through differentiable combinatorial
solvers~\cite{vlastelica2020differentiation} would make suppressing false
bridges learned rather than corrected post hoc. The single
validation-selected threshold makes a related trade: no per-model tuning, but
a model whose score distribution drifts from the validation set is
decided at an operating point chosen for others;
per-model confidence calibration~\cite{guo2017calibration} offers a
threshold-free correction.

\subsection{Scalability and Small-Face Robustness}
\label{sec:limit_facepairing}

\textbf{Quadratic pair domain.}
The scorer evaluates all $N(N{-}1)/2$ unordered pairs and materializes
$N \times N$ score matrices, so memory and computation grow quadratically
with the face count, and assembling the all-shortest-route
context over all pairs adds an $O(N^3)$ term (the all-pairs distances
themselves cost only $O(N^2)$ on the sparse adjacency graph). Within the dataset's 3--250-face range this is
unproblematic on a single consumer GPU, but the cost will bind for industrial
models with many hundreds of faces, where the negative-to-positive imbalance grows
linearly in the face count on top of the quadratic pair cost. Sparse attention offers a remedy:
BigBird~\cite{zaheer2020bigbird} reaches linear complexity with random,
sliding-window, and global attention patterns, and
Exphormer~\cite{shirzad2023exphormer} builds sparse graph-transformer
patterns from expander graphs. Adapted to face pairing, topological adjacency
would define the local windows, and high-confidence seed faces would serve as global tokens.

\begin{figure}[t]
    \centering
    \includegraphics[width=\linewidth]{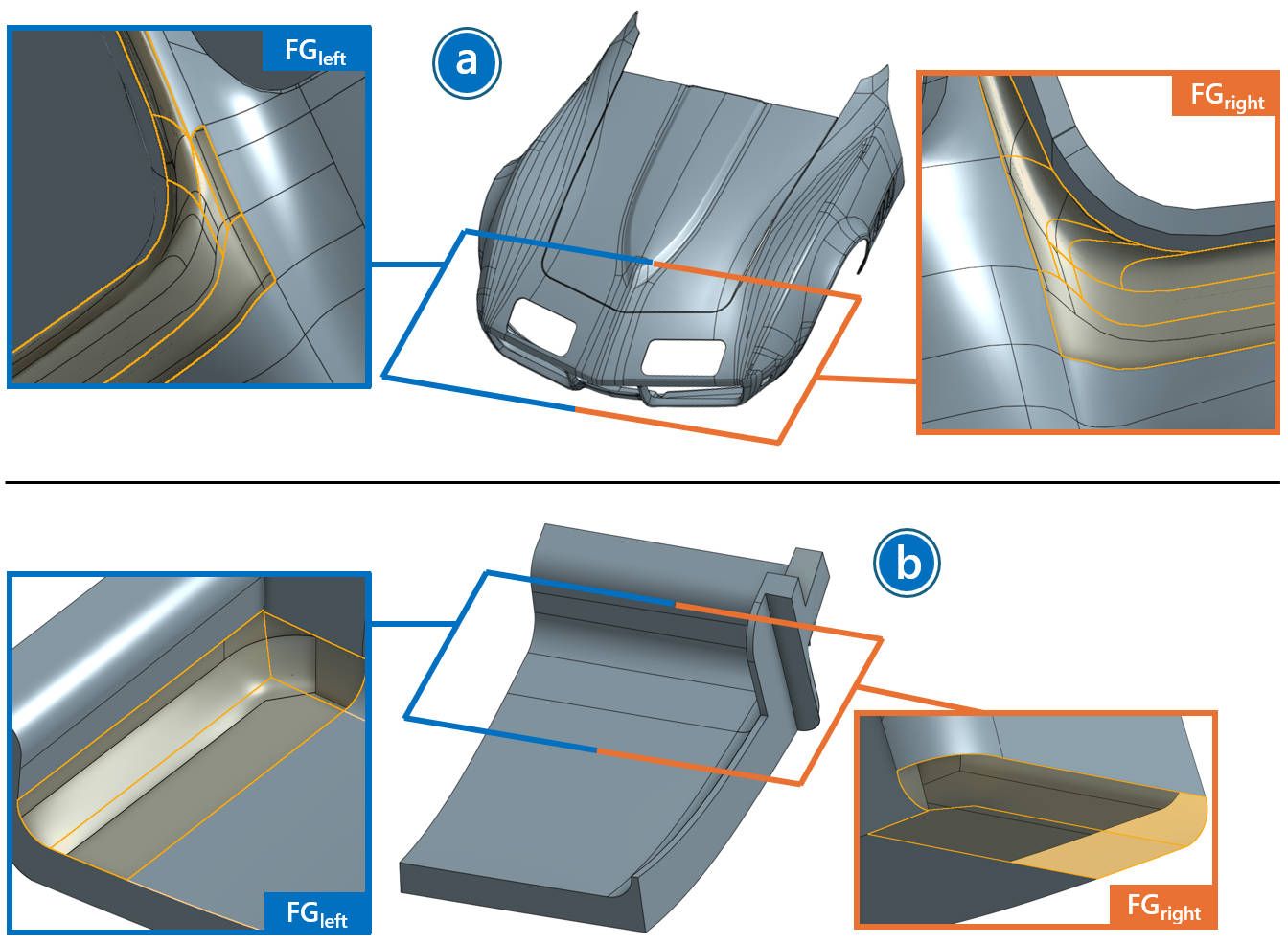}
    \vspace{-6mm}
    \caption{Illustration of two representative failure cases. 
    (a) Small faces arising from filleted transitions lead to unreliable 
    geometric features and ambiguous pairing. (b) Non-thin-wall regions 
    where the face pairing paradigm breaks down due to the absence of 
    well-defined opposing face pairs.}
    \vspace{-5mm}
    \label{fig:limitation}
\end{figure}

\textbf{Small faces.}
Real-world models contain small faces from two sources: split-surface
artifacts, where trimming or Boolean operations subdivide one logical
surface, and intentional features such as fillets, chamfers, and transitional
surfaces. Both are difficult (Fig.~\ref{fig:limitation}-a): small faces
yield unreliable normals and noisy geometric features, their pairing often
admits several valid configurations, and they require disproportionate
annotation effort. Two complementary strategies merit investigation:
\textit{feature suppression}, where established CAD
simplification~\cite{hamdi2012cad,sheffer2001model} removes blend faces and
other details irrelevant to mid-surface topology, and \textit{face
clustering with refitting}~\cite{lai2016small,slyadnev2020simplification},
which merges adjacent small faces of similar geometry, reducing the effective
face count while preserving the essential shape.

\subsection{Beyond Thin Walls}
\label{sec:limit_nonthick}

Face-pairing-based abstraction assumes the input decomposes into thin-walled
regions with well-defined opposing faces. Solid blocks, thick junctions, and
transition masses violate this assumption (Fig.~\ref{fig:limitation}-b), and
the failure takes two forms: \textit{unpairable regions}, where the scorer
correctly finds no partner for some faces and the mid-surface carries holes
or disconnected patches, and \textit{representation inadequacy}, where
forced pairings exist but approximate the underlying geometry poorly and
mislead analysis. Notably,
the first failure is the honest one: the region genuinely has no
opposing-face structure, so the deficiency lies in the paradigm rather than
the prediction. A promising remedy is \textit{hybrid mid-surface modeling}
through virtual decomposition: following Woo et al.~\cite{woo2014divide},
the solid is first split into volumetric regions, for which thick--thin
decomposition provides robust
algorithms~\cite{sun2017decomposing,nolan2019analysis}. Thin-sheet regions
then admit our face-pairing approach directly, while medial-axis-transform
methods~\cite{wang2022mat,kong2024quasi} supply skeletal representations for
the residual thick regions; stitching the two representations at region
boundaries remains the open problem. Alternatives include
erosion-based skeletonization and direct volumetric learning that predicts
mid-surface geometry without explicit face pairing, each trading geometric
fidelity, computational cost, and CAE-integration complexity differently.

\subsection{Annotation Ambiguity and Evaluation Scope}
\label{sec:limit_dataset}

Despite the multi-stage annotation protocol, mid-surface definition is
inherently ambiguous in places: transitional regions between walls of
different thicknesses admit several reasonable pairing interpretations, and
experts may disagree on borderline cases.
This uncertainty propagates into training, where geometrically similar
configurations may receive conflicting supervision. Uncertainty-aware
objectives that model the ambiguity explicitly, or consensus labeling over
several annotators, could soften both effects.

The same ambiguity shapes what the evaluation can measure. Pair metrics
assess the scorer objectively over the complete pair domain, and the
Completion Rate follows the established expert criterion of
MidSurfer~\cite{ye2026midsurfer}---a deliberate choice, because without a
canonical reference mid-surface a fully automated geometric comparison is
ill-posed. Its cost is granularity and scale rather than validity: it
detects topological omissions and additions but neither grades geometric
quality nor extends to arbitrarily large test sets. Automated
topological-equivalence checks against the input solid, together with
group-level metrics for exact variable-cardinality recovery, are the natural
extensions toward larger and finer-grained studies.

\section{Conclusion}
\label{sec:conclusion}

We have presented \toolName{}, to our knowledge the first learning-based
face-pairing method for mid-surface abstraction: a \textit{learned face-pair
scorer} decides every unordered face pair, and a \textit{deterministic
face-group composition} organizes the retained relations into candidate
variable-cardinality $m$-to-$n$ face groups without an additional geometric
threshold. On the MidSurf test set, the scorer reaches an F1-Score of 87.32\%---23.22 points above the strongest rule-based baseline and at least 6.7
points above every learned baseline---and the full pipeline completes
75.42\% of the models, including 61.90\% of the multi-wall category that
the evaluated rule-based implementations do not support. Because the
composition is deterministic, one part yields one result, a property whose
downstream value the FEA case study illustrates.

We have also contributed the MidSurf dataset: 1,575 thin-walled CAD models
with 28,514 manually annotated opposing-face support relations. Looking
ahead, sparse attention for larger industrial models, hybrid modeling beyond
thin walls, group-level metrics, and graph-aware training
(Sec.~\ref{sec:discussion}) mark the immediate next steps.

\section*{References}

\bibliography{sample-base}

\end{document}